\documentclass[fleqn,usenatbib]{mnras}

\usepackage{newtxtext,newtxmath}

\usepackage[T1]{fontenc}
\usepackage{comment}

\DeclareRobustCommand{\VAN}[3]{#2}
\let\VANthebibliography\thebibliography
\def\thebibliography{\DeclareRobustCommand{\VAN}[3]{##3}\VANthebibliography}

\usepackage{graphicx}	
\usepackage{amsmath}	

\newcommand{\angstrom}{\textup{\AA}}
\usepackage{float}
\usepackage{subcaption}
\usepackage{cleveref}
\usepackage{upgreek}
\usepackage{float}
\usepackage[toc,title,page]{appendix}
\usepackage[section]{placeins} 

\usepackage{xspace}
\newcommand{\ppxf}{\textsc{ppxf}\xspace}

\title[Star formation inside galactic outflows]{Signatures of star formation inside galactic outflows}

\author[Dily Duan Yi Ong et al.]{Dily Duan Yi Ong,$^{1,2,3}$\thanks{E-mail: dlo26@cam.ac.uk}
 Francesco D'Eugenio,$^{1,2}$
Roberto Maiolino,$^{1,2,4}$
Santiago Arribas,$^{5}$
\newauthor Francesco Belfiore,$^{6,7}$
Enrica Bellocchi,$^{8,9}$
Stefano Carniani,$^{10}$
Sara Cazzoli,$^{11}$
Giovanni Cresci,$^{7}$
\newauthor Andrew Fabian,$^{3}$
Wako Ishibashi,$^{12,13}$
Filippo Mannucci,$^{7}$
Alessandro Marconi,$^{14,7}$
Helen Russell,$^{15}$
\newauthor Eckhard Sturm$^{16}$ and
Giacomo Venturi$^{10}$
\\
$^{1}$Cavendish Laboratory, University of Cambridge, 19 J. J. Thomson Ave., Cambridge CB3 0HE, UK\\
$^{2}$Kavli Institute for Cosmology, University of Cambridge, Madingley Road, Cambridge CB3 0HA, UK\\
$^{3}$Institute of Astronomy, University of Cambridge, Madingley Road, Cambridge, CB3 0HA, UK\\
$^{4}$Department of Physics and Astronomy, University College London, Gower Street, London WC1E 6BT, UK\\
$^{5}$Centro de Astrobiolog\'ia (CAB), CSIC-INTA, Cra. de Ajalvir, km 4, E-28850, Torrej\'on de Ardoz, Madrid, Spain\\
$^{6}$European Southern Observatory, Karl-Schwarzschild Stra\ss e 2, D85748 Garching bei M\"unchen, Germany\\
$^{7}$INAF-- Arcetri Astrophysical Observatory, Largo E. Fermi 5, I50125, Florence, Italy\\
$^8$ Departamento de F\'isica de la Tierra y Astrof\'isica, Fac. CC. F\'isicas, Universidad Complutense de Madrid, Plaza de las Ciencias, 1, Madrid 28040, Spain\\
$^9$ Instituto de F\'isica de Part\'iculas y del Cosmos (IPARCOS), Fac. CC F\'isicas, Universidad Complutense de Madrid, E-28040 Madrid, Spain\\
$^{10}$Scuola Normale Superiore, Piazza dei Cavalieri 7, I-56126 Pisa, Italy\\
$^{11}$Instituto de Astrof\'isica de Andaluc\'ia (IAA-CSIC), Glorieta de la Astronom\'ia s/n, E-18008, Granada, Spain\\
$^{12}$Physics Institute, University of Zurich, Winterthurerstrasse 190, CH-8057 Z\"urich, Switzerland\\
$^{13}$Istituto Ricerche Solari (IRSOL), Universita della Svizzera italiana (USI), CH-6605 Locarno Monti, Switzerland\\
$^{14}$Dipartimento di Fisica e Astronomia, Universit\`a degli Studi di Firenze, Via G. Sansone 1, I-50019 Sesto Fiorentino, Firenze, Italy\\
$^{15}$School of Physics and Astronomy, University of Nottingham, University Park, Nottingham NG7 2RD, UK\\
$^{16}$Max-Planck-Institut f\"ur extraterrestrische Physik, Giessenbachstrasse 1, D-85748 Garching, Germany
}

\date{Accepted XXX. Received YYY; in original form ZZZ}

\pubyear{2025}

\begin{document}
\label{firstpage}
\pagerange{\pageref{firstpage}--\pageref{lastpage}}
\maketitle

\begin{abstract}
Observations have suggested that galactic outflows contain substantial amounts of dense and clumpy molecular gas, creating favourable conditions for igniting star formation. Indeed, theoretical models and hydrodynamical simulations have suggested that stars could form within galactic outflows, representing a new mode of star-formation that differs significantly from the typical star formation in star forming discs. In this paper, we examine 12 local galaxies with powerful Active Galactic Nuclei and high star-formation rate using spectroscopic data from the X-shooter spectrograph at the Very Large Telescope. We investigate the excitation mechanism and physical properties of these outflows via spatially resolved  diagnostic diagrams (along with tests to rule out contribution by shocks and external photoionisation). Out of the seven galaxies with clearly detected outflows, we find robust evidence for star formation within the outflow of one galaxy (IRAS 20551-4250), with two additional galaxies showing tentative signs (IRAS 13120-5453 and F13229-2934). Therefore, our findings support previous results that star formation inside outflows can be a relatively common phenomenon among these active galaxies and may have played an important role in the formation and evolution of the spheroidal component of galaxies.

\end{abstract}




\section{Introduction}
Galactic outflows are commonly observed in active galaxies. They can be driven either from accreting supermassive black holes
(i.e., active galactic nuclei, AGN), or by supernova explosions and radiation pressure from massive young stars (starburst-driven). Galactic outflows may have a significant influence on their host galaxies, typically via a negative feedback effect on star formation. This effect is obtained via gas ejection (thus removing the material needed for forming new stars) and because outflows can heat up the interstellar medium (ISM), so fragmentation and gravitational collapse are less likely to take place~\citep{granato2004,king2010,fabian2012,kingpounds2015}. However, some recent theoretical predictions and observational evidence are also in favour of scenarios where star formation can be induced by outflows impacting the galactic discs or circumgalactic clouds
\citep{ishi2012, 2013MNRAS.431..793Z, silk2013, 2014MNRAS.439..400Z, silk2010, nayazu2012, dugan2014, mukherjee2018, gaibler2012}.
In this positive feedback scenario high-velocity galactic outflows can compress molecular gas within the ISM and circumgalactic medium (CGM), leading to the fragmentation and gravitational collapse of molecular clouds and potentially igniting star formation within galactic discs~\citep{rees1989,nayazu2012,ishi2012,zu2013,bieri2016}.
Evidence of SF induced by outflows and jets has been found in a number of cases~\citep{santoro2016,crockett2012,croft2006,cresci2015,elbaz2009,salome2015,lacy2017,molnar2017}.

Moreover, observations have shown evidence of galactic outflows containing a significant quantity of cold molecular gas~\citep{feruglio2010,sturm2011,cicone2014,Combes2014,Sakamoto2014,Garcia2015,Fluetsch2019,Fluetsch2021} with a remarkably large proportion of dense molecular gas~\citep{aalto2012,aalto2015,walter2017,privon2017,cicone2020}, which is highly clumpy in nature~\citep{pereira2016,finn2014,borguet2012,feruglio2015}. These observational findings suggest that gas in galactic outflows has the physical conditions for forming stars inside the outflow. Indeed, a number of models have proposed a scenario in which gas in galactic outflows may undergo compression, gravitational collapse and form stars {\it inside} the outflow~\citep{gaibler2012,ishi2012,ishibashifabian2014,ishi2013,silk2013,zubovas2013b,dugan2014,zubovasking2014,mukherjee2018,wangloeb2018}. This phenomenon has been found to happen in hydrodynamical zoom-in simulations~\citep{yu2020,elbadry2016}.

This potential new mode of star formation {\it inside} galactic outflows differs significantly not only from the typical star formation within galactic discs, but also from the star formation resulting from the ``standard'' positive feedback scenario described earlier, i.e. star formation triggered by the compression of gas caused by the interaction with outflows/jets. Stars formed in galactic outflows are born with high radial velocities on nearly radial orbits, and, depending on their radius and velocity at the time of formation, they either end up escaping the galaxy and/or the halo, or becoming gravitationally bound~\citep{zu2013}. In the latter case, they can potentially make a significant contribution to the formation and evolution of the spheroidal component of galaxies, i.e. bulge and halo, and even to the formation of elliptical galaxies~\citep{ishi2013,dugan2014,yu2020}. Some models suggested that this star formation mode can reach a few/several 100 M\textsubscript{\(\odot\)} yr$^{-1}$ ~\citep{ishi2012,silk2013}. Another implication of this star formation mode is its ability to enrich the chemical composition of the circumgalactic and intergalactic medium through supernova explosions of young stars on large orbits~\citep{maiolino2017}; this is especially true when taking into account that outflowing gas is generally more metal enriched than the ISM in the disc~\citep{chisholm2018}. Moreover, certain models indicate that it could play a substantial role in driving the observed star-formation rate in high-redshift galaxies~\citep{silk2013}. The same scenario could explain the discovery of metal-rich stars in the MW  with highly radial velocities~\citep{belokurov2018}.

The first clear observational evidence for star formation occurring inside an outflow was found by~\citet{maiolino2017}. They analysed the spectroscopic data obtained with the X-shooter, SINFONI, and the multi-unit spectroscopic explorer (MUSE) instruments on the Very Large Telescope (VLT) of the  system IRAS F23128-5919, characterized by a prominent galactic outflow~\citep{leslie2014,Bellocchi2013,arribas2014,cazzoli2016,piqueras2012}. The datasets presented unambiguous evidence of star formation occurring within the galactic outflow using spatially resolved BPT diagnostics~\citep{baldwin1981}, near-IR diagnostics and kinematics of the young stellar populations. \citet{gallagher2019} used a similar approach but looked into a much larger galaxy sample from the MaNGA-SDSS4 DR2 survey~\citep{bundy2015}. By analysing the integral field spectra of 2,800 galaxies, they identified a subsample of 37 galaxies with clear outflows, and in which one third shows evidence for star formation, and approximately half of the outflows display indications of at least some level of star formation. This study reveals that this new mode of star formation within galactic outflows is common among galaxies. \citet{2019MNRAS.486..344R} and~\citet{perna2021} also report similar findings using similar datasets. More recently, clear evidence for star formation happening inside the outflow of the nearby, prototype starburst galaxy M82 has been presented by~\citet{rao2025}.

This mode of star formation may have not been widely recognised in the past because the star formation diagnostics are easily dominated by the presence of even faint AGN and shocks~\citep{gallagher2019}. Fortunately, with the unprecedented improvements in sensitivity of spectrographs, such as the X-shooter and MUSE, high quality and also spatially resolved spectra spanning a wide wavelength range are available and even faint signals can be detected.

In this paper we report the detailed analysis of the X-shooter spectra of 7 luminous and ultra luminous infrared galaxies (i.e., U/LIRGs) with evidence for outflows. 
This study used a similar analytical approach and diagnostics as~\citet{maiolino2017} and~\citet{gallagher2019}, but differs in a number of key aspects. High quality data were obtained using the X-shooter (also used in~\citealt{maiolino2017}, but for one galaxy only), which operates at a much higher spectral resolution ($R$ = 3000 - 18000, depending on wavelength and slit width), compared with MUSE ($R$ $\sim$ 3000), and also higher sensitivity compared with MaNGA ($R$ $\sim$ 2000, used by \citealt{gallagher2019} and by \citealt{2019MNRAS.486..344R}). This enables the measurement of previously undetectable, faint signals, especially for the weak galactic outflows. This is crucial not only in identifying and disentangling galactic outflow components from the galaxy spectra (discussed in Section \ref{sec:fitting}), but also increase accuracy in determining the values of flux of emission lines, which are then used in diagnostic diagrams~\citep{baldwin1981,Kewley2001,kewley2006}. In addition, the galaxy sample was selected to cover more extreme cases, i.e. with higher star formation rate and potentially more powerful AGN, as these are more likely  to drive powerful galactic outflows.

We find robust evidence for star formation within the outflow of one galaxy (IRAS 20551-4250), with two additional galaxies showing tentative signs (IRAS 13120-5453 and F13229-2934), confirming previous findings that star formation inside outflows may be relatively common.

Throughout this paper, we assume a flat $\Lambda$CDM cosmology with parameters from the Planck 2018 results ($H_0 = 67.66$~km~s$^{-1}$~Mpc$^{-1}$, $\Omega_m = 0.3111$; \citealt{planck2020}).

\section{Methodology}
\begin{table*}
\begin{tabular}{lllllll}
\hline
Galaxy name         & Redshift $z$ & Scale & $\log(L_{\rm IR}/L_\odot)$ & Seeing (") & RA (J2000) & DEC (J2000) \\
                    &              & (kpc/") &  &  &  &  \\
\hline
IRAS 20551-4250     & 0.0430       & 0.877 & 12.23 & 0.873 & 20:58:26.82 & $-$42:38:59.41 \\
IRAS 13120-5453     & 0.031249     & 0.646 & 12.26 & 0.803 & 13:15:06.3 & $-$55:09:23 \\
F13229-2934/NGC5135 & 0.0137 & 0.280 & 11.29 & 0.745 & 13:25:44.0 & $-$29:50:01 \\
F22491-1808         & 0.07775      & 1.521 & 12.21 & 1.18 & 22:51:49.20 & $-$17:52:21.76 \\
NVSSJ151402+015737  & 0.09537      & 1.828 & 12.11 & 0.757 & 15:14:02 & $+$01:57:37 \\
LEDA 166649         & 0.0629       & 1.252 & 11.39 & 0.855 & 19:56:35.73 & $+$11:19:05.81 \\
NGC 7130            & 0.016151 & 0.329 & 11.41 & 0.645 & 21:48:19.5 & $-$34:57:05 \\
\hline
F00509+1225         & 0.061169     & 1.220 & 11.52 & 0.652 & 00:53:34.93 & $+$12:41:35.93 \\
F21130-4446         & 0.092554     & 1.780 & 11.83 & 0.945 & 21:16:18.58 & $-$44:33:37.14 \\
F17207-0014         & 0.013754     & 0.291 & 12.18 & 0.64  & 17:23:21.96 & $-$00:17:00.81 \\
IC 4518B            & 0.016568     & 0.320 & 11.11 & 0.69  & 14:57:43.1 & $-$43:07:48 \\
F23389+0300         & 0.145        & 2.626 & 11.97 & 0.67  & 23:41:30.31 & $+$03:17:26.57 \\
\hline
\end{tabular}
\caption{Properties of 12 galaxies observed by the X-shooter in the ESO programme 0103.B-0478(A) and 097.B-0918(A). Columns show: redshift, physical scale in kpc per arcsecond, logarithm of the infrared luminosity ($L_{\rm IR}=L(8-1000\,\mu$m) in units of solar luminosity), average seeing during observations, and right ascension and declination (J2000). Coordinates, scales, and infrared luminosities for F13229-2934/NGC5135, NGC 7130, and IC 4518B are from~\citet{arribas2008}. Coordinates for IRAS 20551-4250 and infrared luminosities for IRAS 20551-4250 and IRAS 13120-5453 are from SIMBAD and NED. Infrared luminosities for F22491-1808, F00509+1225, and F17207-0014 are from~\citet{2003AJ....126.1607S}. Infrared luminosities for LEDA 166649, F21130-4446, and F23389+0300 are from NED. Infrared luminosity for NVSSJ151402+015737 (also known as IRAS F15116+0209) is estimated from WISE W4 band photometry. Physical scales are calculated from redshifts using Planck 2018 cosmology for galaxies not in~\citet{arribas2008}. Coordinates for remaining galaxies are extracted from X-shooter FITS file headers. We could not thoroughly perform analysis and reach a conclusion on the last 5 galaxies in the table because of: (i) poor SNR due to limitations imposed by visibility or exposure time during observation (galaxy F17207-0014, IC 4518B and F23389+0300), (ii) outflow is not detected (galaxy F21130-4446) and (iii) stellar continuum cannot be fitted and caused difficulties in disentangling the galactic outflow components from the stellar continuum (stellar continuum of galaxy F00509+1225 is dominated by the broad-line region). Outflow disentanglement is discussed in Section~\ref{sec:fitting}.}
\label{tab:my-table}
\end{table*}
\subsection{Sample selection, observations and data processing}

We used spectroscopic data obtained by the X-shooter on the VLT in the ESO programme 0103.B-0478(A) and 097.B-0918(A)\footnote{The associated raw data are
available at the ESO Science Archive Facility: \url{http://archive.eso.org/cms.html}.}. The sample consisted of 12 ULIRGs and LIRGs selected from the parent sample obtained by~\citet{arribas2008}, who observed 42 LIRGs and ULIRGs with the VLT-VIMOS integral field spectroscopic mode. The subsample of 12 galaxies was selected to show evidence of outflow in the VIMOS spectra. Out of the sample of
12 observed galaxies only 7 had the quality required by our analysis, and only these were studied further. Indeed, out of the 5 remaining galaxy spectra, 3 of them (F17207-0014, IC 4518B and F23389+0300) have very low signal-to-noise ratio (SNR) due to limitations imposed by visibility and exposure time during observation. Another galaxy, F00509+1225, is dominated by the broad-line region, which complicates the fitting of the stellar continuum, and hence caused difficulties in disentangling the galactic outflow components from the stellar continuum (outflow disentanglement is discussed in Section \ref{sec:fitting}). In the last of the five galaxies, F21130-4446, no outflow is detected.

X-shooter covers, in a single exposure, the spectral range from 3000$\angstrom$ to 24800$\angstrom$, enabling simultaneous access to fundamental gas emission lines needed for investigating excitation mechanism and physical properties of the ionised gas, including but not limited to: H$\alpha$, H$\beta$, [OII]$\lambda\lambda3726,3729\angstrom$, [OIII]$\lambda\lambda4959,5007\angstrom$, [OI]$\lambda\lambda6300,6364\angstrom$, [NII]$\lambda\lambda6548,6584\angstrom$ and [SII]$\lambda\lambda6717,6731\angstrom$. X-shooter operates at intermediate spectral resolution ($R$ = 3000 - 18000, depending on wavelength and slit width) with fixed \'echelle spectral format (prism cross-dispersers) in three spectroscopic arms, each with optimised optics, dispersive elements. The three spectral regions are named UVB ($\lambda = 3000-5595 \angstrom$), VIS ($\lambda = 5595-10240 \angstrom$) and NIR ($\lambda = 10240-24800 \angstrom$). Unlike~\citet{maiolino2017}, only UVB and VIS spectroscopic data are used in our analysis because our NIR data are too noisy for our spectral decomposition study. The seeing during the observations was between 0.64" to 1.18" (see~\Cref{tab:my-table}). Observations were executed in the fixed offset to sky mode. Fixed offset to sky mode is used for observations of extended objects, such as galaxies, where there is no or not enough pure sky along the slit for a good sky subtraction. It involves the alternation between an object and sky position with a possible addition of a small jittering around the object and the sky position,
which allows a good sky background subtraction\footnote{For more information, please refer to the X-shooter User Manuals available at \url{https://www.eso.org/sci/facilities/paranal/instruments/xshooter/doc.html}.}.
The raw X-shooter data were reduced and calibrated using the X-shooter data reduction pipeline\footnote{available at \url{http://www.eso.org/sci/software/pipelines/}}. Twelve spectra of telluric standard stars were taken at regular intervals between galaxy observations.
These stars were used to correct for atmospheric absorption features in the VIS range of the galaxy spectra in cases where these features are at the same wavelength as the emission lines of interest. Before the analysis, the spectra were brought close to the rest-frame wavelength using redshift values from the literature before fitting with \ppxf.

\subsection{Spectral fitting and outflow disentanglement}\label{sec:fitting}

We extracted galaxy spectra along the slit at different offsets from the galaxy centre, and with increasing apertures (from 0.5" to 2.5"). There is a trade-off between spatial information and the SNR. The minimum aperture size is set to be equal to the seeing value stated in~\Cref{tab:my-table} during observation for each galaxy, and then it is gradually increased as the SNR decreases along the slit, i.e. further away from the central region of galaxies.

\subsubsection{Fitting and deblending stellar continua}

Galaxy spectra were analysed by fitting the stellar continuum and the multiple emission line kinematic components using an adapted version of the \ppxf routine~\citep{Cappellari2004,Cappellari2017,Cappellari2022}. As initial noise estimate, we used the values reported by the data reduction pipeline. The reduced chi-square per degree of freedom in the \ppxf output needs to be $\sim 1$ for the errors of flux of emission lines to be properly estimated. In most cases, \ppxf had to be re-run with the input noise array re-scaled by multiplying it by the square root of the previous reduced chi-square value. Although binning of flux can increase the SNR, we did not perform that to avoid mixing regions of vastly different properties and losing spatial information. 

To model the stellar continuum, we used a set of simple stellar-population spectra obtained by combining a high-resolution version of the C3K stellar spectral libraries~\citep[$R=10,000$;][]{Conroy2019} with MIST isochrones \citep{Choi2016}. This choice ensures the full spectral coverage and sufficient spectral resolution to match the X-shooter data \footnote{These spectral templates are available from the corresponding author, C. Conroy, upon reasonable request.}. The stellar continuum was fitted, i.e. disentangled from the nebular emission lines, with a total of 1248 stellar templates, with the assumption of a non-Gaussian Line-Of-Sight Velocity Distribution (LOSVD), parametrised as a 4\textsuperscript{th}-order Gauss-Hermite series, with moments $v$, $\sigma_{\mathrm{v}}$, $h_3$, and $h_4$. We further used a 10\textsuperscript{th}-order multiplicative Legendre polynomial, for taking into account the combined effect of dust reddening, flat fielding, and other calibration issues. The order of the polynomial was required to obtain a proper fit of the very broad spectral range of the continuum. We used no additive Legendre polynomial. An additive polynomial would be useful to correct for the EW of the stellar absorption lines in the case that the templates are not adequate, however our fit are always reproducing well the stellar features and anyway we are not interested in measuring the EW of the stellar population features. Ideally, the stellar fit should permit different velocities for the distinct stellar populations. Specifically, it should be possible to assign a distinct velocity to the old stellar population relative to the young stellar population. Unfortunately, implementing this in practice is challenging due the additional degrees of freedom for different velocities of distinct stellar populations, which would lead to a significant increase in degeneracy, making it more difficult to deblend the stellar features from the nebular emission lines and determine the unique velocity of individual gas components.
Prior to the fit, we convolve all templates with the line-spread function of X-shooter data.

\begin{figure*}
	\centering
	\begin{subfigure}{\textwidth}
        \centering
		\includegraphics[width=\textwidth]{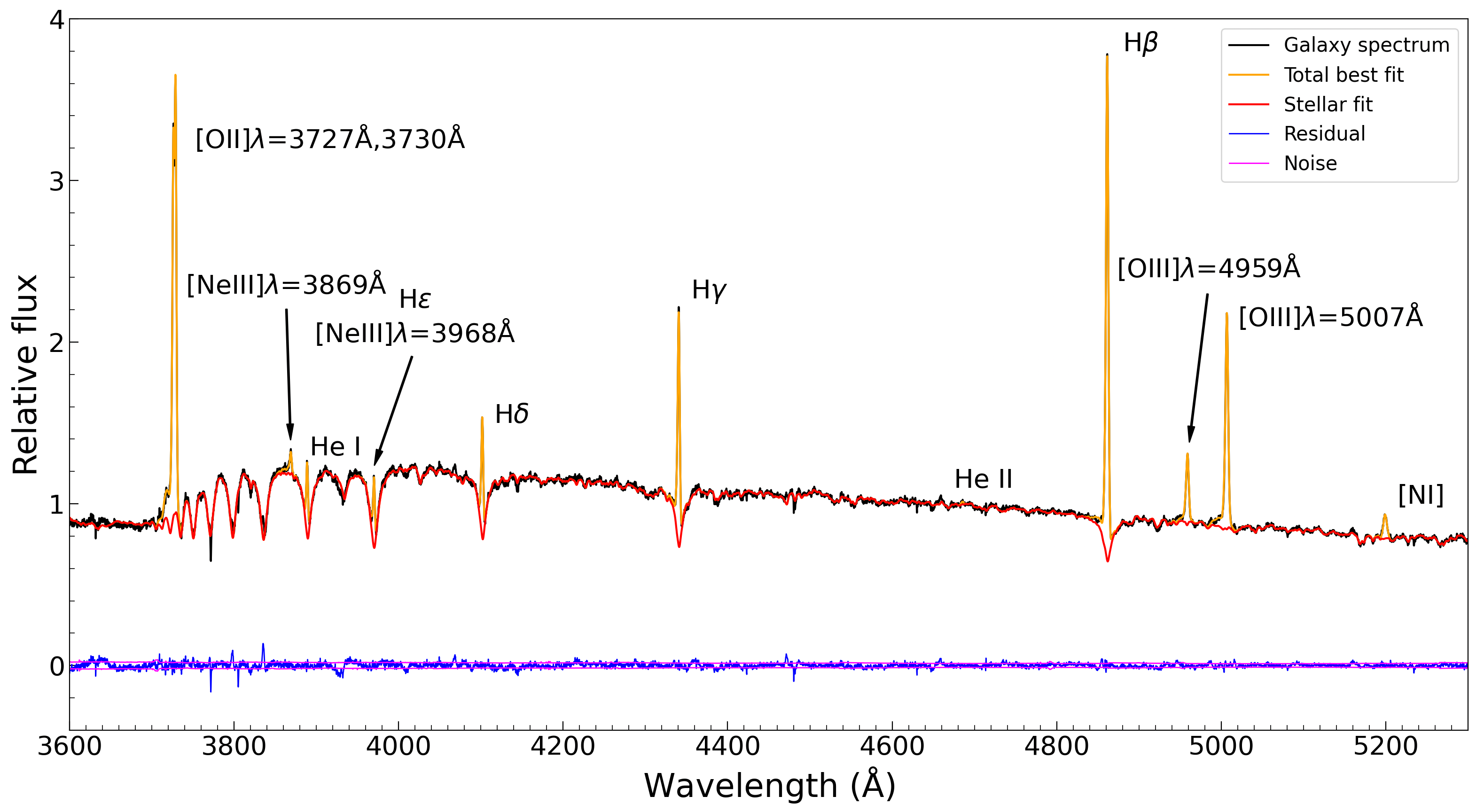}
		\caption{} 
		\label{circle}
	\end{subfigure}
	\begin{subfigure}{\textwidth}
        \centering
		\includegraphics[width=\textwidth]{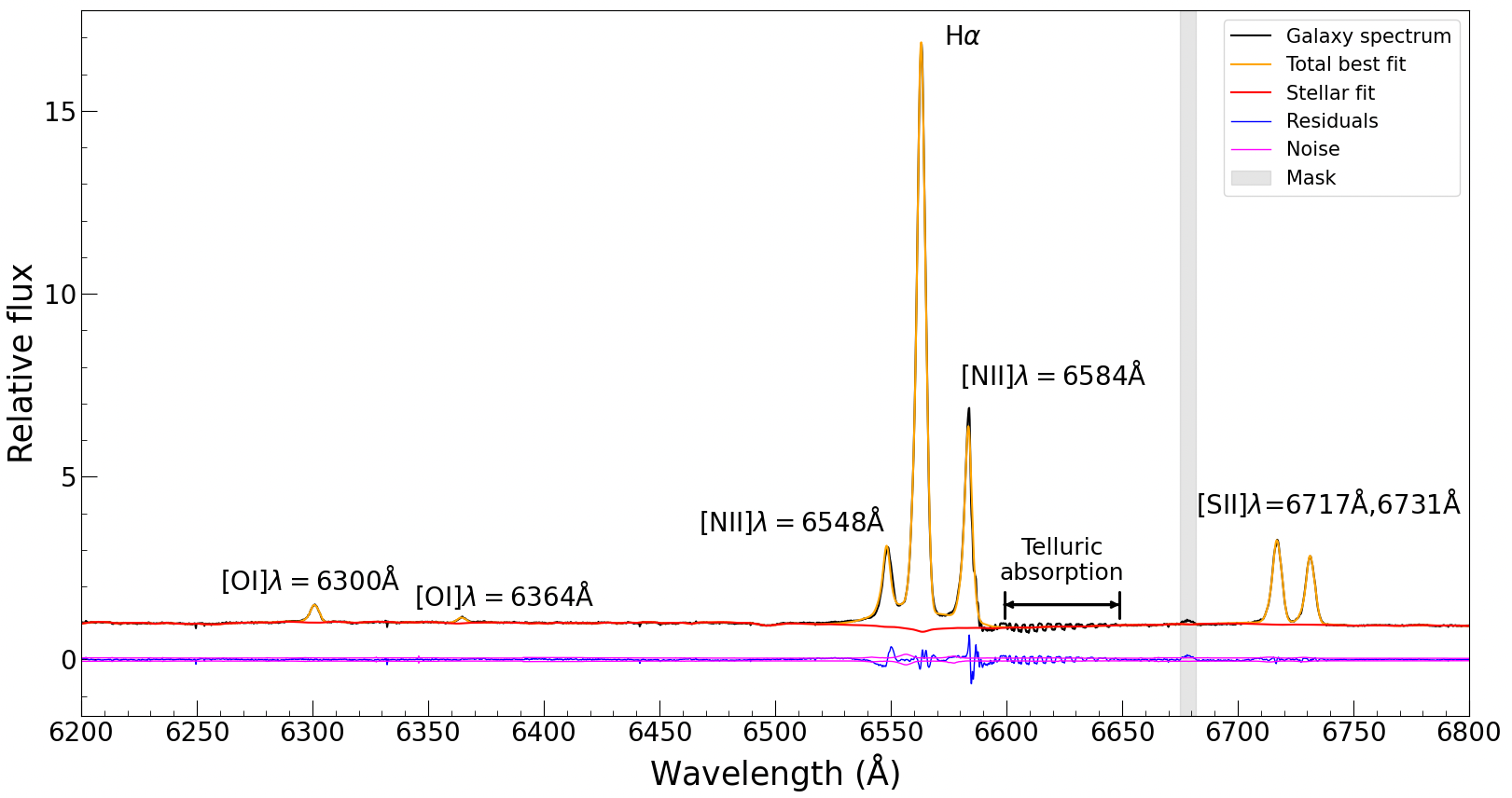}
		\caption{} 
		\label{Mean}
	\end{subfigure}
	\caption{An example of simultaneously fitting the stellar continuum (red) and emission lines (orange) of the spectrum (black) of galaxy (IRAS 20551-4250) extracted from the central region in the (a) UVB range and (b) VIS range. Shown below the fit are the residuals (blue) from the PPXF fitting and the noise spectrum (magenta) from the observation. Emission lines (He I line in this example) unrelated to the analysis has been masked (grey vertical strip) to avoid interrupting the stellar continuum fitting.} 
    \label{fig:IRAS_20551-4250_whole}
\end{figure*}

\begin{figure*}
    \includegraphics[width=2\columnwidth]{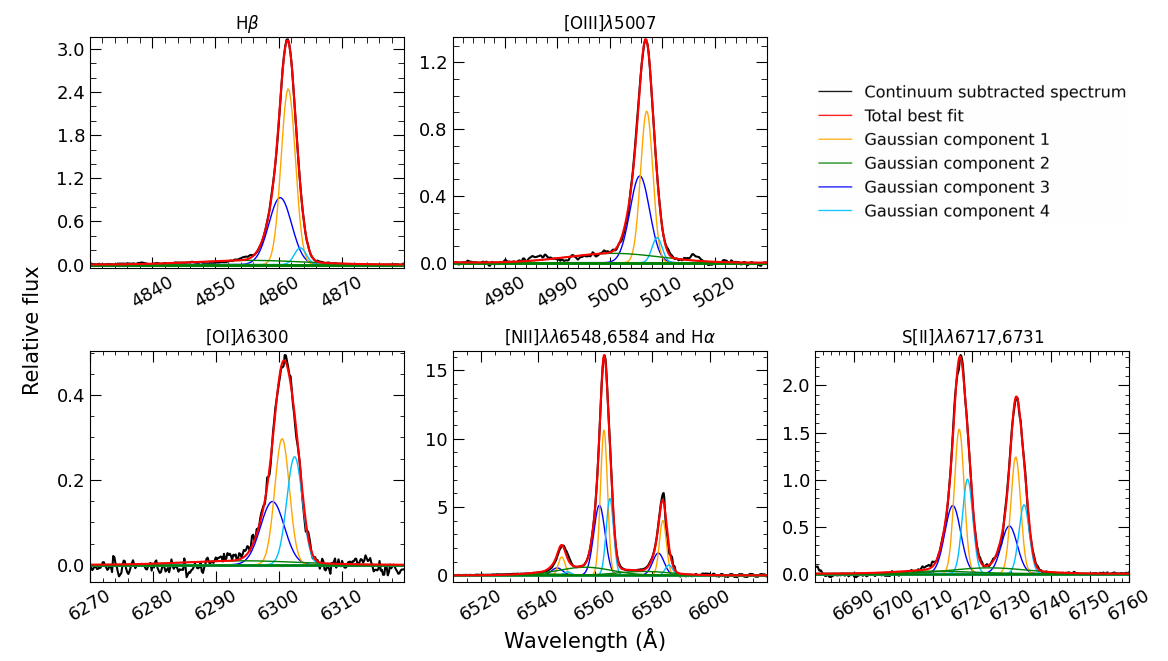}
    \caption{Subsections of continuum-subtracted X-shooter spectra (black) of the galaxy IRAS 20551-4250, extracted from the central region around the relevant emission lines for BPT-diagnostics, displaying the decomposition by 4 Gaussian components (see legend) representing the narrow components associated with the gas in the galactic disk and the broad components tracing the outflows. For clarity of presentation, the best-fit stellar continuum is subtracted in this figure. However, the full spectral fit with \texttt{ppxf} was performed on the stellar continuum and emission lines simultaneously.}
    \label{fig:IRAS_20551-4250_zoom}
\end{figure*}

\subsubsection{Fitting nebular emission lines and disentangling outflows}

The emission lines were fitted simultaneously with the stellar continuum. We note that while some figures in this paper show continuum-subtracted spectra for clearer visualisation of the emission line components, the actual spectral fitting was always performed with both the stellar continuum and emission lines fitted jointly. Initially a single Gaussian per emission line was used.
To model the nebular emission lines, we used a modified \ppxf code to simultaneously fit multiple Gaussian components, with the velocity ($v$) and velocity dispersion ($\sigma_v$) tied across all lines for a given component. The optimal number of components was found by balancing two competing requirements in an iterative process. First, we established a minimum goodness-of-fit by adding components until the structured residuals were minimised; the fit was considered acceptable once the residual peak was below 3 times the noise or less than 3\% of the total line flux. Second, to prevent over-fitting, we imposed a maximum complexity limit defined by a degeneracy criterion. If adding a component caused any two Gaussians to converge on nearly identical parameters ($|v_i - v_j| < 10$~km~s$^{-1}$ and $|\sigma_i - \sigma_j| < 10$~km~s$^{-1}$), the model was considered degenerate and the previous, ($n-1$)-component solution was adopted. The final multi-component fit is a superposition of narrow components from the galactic disc and broad, blueshifted components from outflows~\citep{maiolino2017,gallagher2019}. We define an outflow kinematically as a component with a blueshift $>$50~km/s and FWHM$>$200~km/s, corresponding to an outflow velocity $|v|+2\sigma > 200$~km/s~\citep{maiolino2017}. We note that the receding (redshifted) outflow components are typically absent due to dust attenuation by the galactic disc.

Key nebular lines for subsequent analysis including H$\alpha$, H$\beta$, [OII]$\lambda\lambda3727, 3730 \angstrom$, [OIII]$\lambda\lambda4959,5007\angstrom$, [OI]$\lambda\lambda6300,6364\angstrom$, [NII]$\lambda\lambda6548,6584\angstrom$ and [SII]$\lambda\lambda6717,6731\angstrom$ were fitted in all 7 galaxy spectra. Other weaker emission lines, including H$\gamma$, H$\delta$, H$\varepsilon$, He I, He II, [NeIII]$\lambda\lambda3869,3968\angstrom$ and [NI] were fitted as well where possible, depending on whether the available wavelength range includes those lines and if the SNR is high enough. In cases where these lines are too weak to be detected or decomposed by \ppxf, we masked them with a mask width $w=2 \times FWHM$, to prevent their signals from biasing the stellar continuum fitting. The line flux ratios of the density-sensitive doublets [OII]$\lambda\lambda3727, 3730 \angstrom$ and [SII]$\lambda\lambda6717,6731\angstrom$ were constrained within the ranges 0.35--1.5 and 0.44--1.44, respectively, as allowed by atomic physics~\citep{osterbrock2006}. For the doublets [OIII]$\lambda\lambda4959,5007\angstrom$, [OI]$\lambda\lambda6300,6364\angstrom$, [NII]$\lambda\lambda6548,6584\angstrom$ and [NeIII]$\lambda\lambda3869,3968\angstrom$, their flux ratios were fixed to 2.99, 3.1, 2.96, and 3.3, respectively, according to their Einstein coefficients~\citep{osterbrock2006}.
We report three-$\sigma$ upper limits on the flux of undetected emission lines, using the error estimated by \ppxf.

\subsubsection{Fitting spectra across the UVB and VIS range}
As mentioned, emission lines of the same gas component were required to share the same kinematics parameters (velocity $v$ and dispersion of velocity $\sigma_{v}$) under the assumption that originated from the same emitting medium. Their flux amplitudes are allowed to vary individually. In theory, the same gas component in the UVB and VIS spectrum should also share the same $v$ and $\sigma_{v}$. But in reality, slightly different regions of the galaxy were sampled due to the difference in slit width used in the UVB ($1"$, $R=5400$) and VIS ($0.9"$, $R=8900$) arm of X-shooter. Therefore, the $v$ and $\sigma_{v}$ of the same gas component in the UVB and VIS spectra were fitted separately and allowed to vary within $\pm$20~percent\footnote{This percentage value has to be small enough to ensure the same gas component is being traced, while large enough to accommodate the discrepancy that arises from using different spectral resolutions for the UVB and VIS range during observations. We have tested a range of percentage from $\pm5$ to $\pm30$~percent and in most cases, $\pm20$~percent strikes the best balance.} of each other.
One of the two spectra is first fitted freely, i.e. without specific bounds to the output parameters, and its solution will be used as the dominant solution. Then, the other spectrum is fitted with the output parameters bounded within $\pm$20~percent of the dominant solution. In most cases, the UVB part of the spectrum is fitted first because (i) it has emission lines that are more spectrally separated, unlike the VIS part, where H$\alpha$ and [NII] doublet lines are usually highly blended, and (ii) it contains more and stronger stellar features, which help constrain the stellar continuum fitting.

\subsubsection{Star formation rate calculation}
\label{sec:sfr_calc}

We derive star formation rates (SFRs) for the outflow components whose line ratios, as determined from the BPT diagram~\citep{baldwin1981} (Section~\ref{section:bpt}), are consistent with ionisation from young, massive stars. The SFR is calculated from the H$\alpha$ luminosity, L(H$\alpha$), using the calibration from~\citet{kennicutt2012}, which assumes a~\citet{chabrier2003} initial mass function (IMF):
\begin{equation}
\label{eq:sfr}
\text{SFR}~[\text{M}_{\odot}~\text{yr}^{-1}] = \text{L}(\text{H}\alpha)~[\text{erg s}^{-1}] \times 5.5 \times 10^{-42}.
\end{equation}
The H$\alpha$ fluxes for each kinematic component are obtained directly from our \texttt{ppxf} spectral decomposition in absolute flux units. The H$\alpha$ luminosity for each kinematic component is then calculated from the absolute flux using the luminosity distance, $D_L$, determined from each galaxy's redshift. A key consideration is that the H$\alpha$ emission used to measure the star formation rate may be contaminated by other ionising sources. To mitigate this, we carefully assess these contributions: we use BPT diagrams to exclude AGN contamination (Section~\ref{section:bpt}), explore the possibility of external gas photoionisation (Section~\ref{section:photoion}), and analyse shock-sensitive line ratios (Section~\ref{section:shock}). Furthermore, we do not correct the H$\alpha$ fluxes for dust extinction, as reliable Balmer decrement measurements (H$\alpha$/H$\beta$) are precluded by the different slit widths used for the UVB and VIS arms, which result in slightly different spatial regions being sampled. Therefore, all derived SFRs should be considered as conservative lower limits.

\section{Results}\label{sec:results}
\subsection{Spatially resolved BPT-diagnostics}\label{section:bpt}
\begin{figure*}
	\includegraphics[width=2\columnwidth]{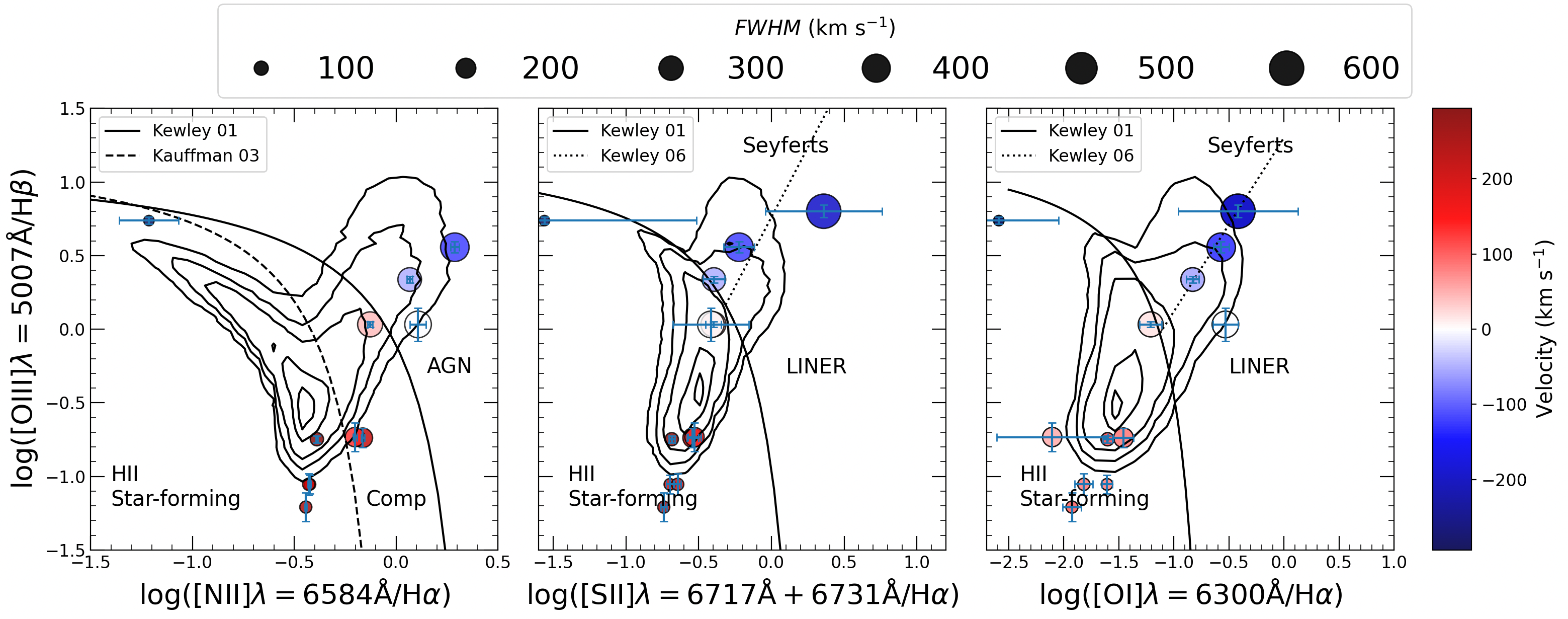}
    \caption{A typical case (4 out of 7 galaxies analysed) where outflows are in the AGN excitation region. The flux ratios of the individual components extracted at different apertures for galaxy NVSSJ151402+015737 are plotted on the [NII] (left), [SII] (middle) and [OI] (right) diagnostic diagrams. The velocity of each component is colour-coded, from red (more redshifted) to blue (more blueshifted). The symbols are proportional to the {\it FWHM} of the component, as indicated by the legend above.  The black line contours indicate the distribution of galaxies and AGN (10, 30, 68, 85 and 95~percent) from the SDSS.}    
    \label{fig:NVSSJ151402+015737_bpt}
\end{figure*}

\begin{figure*}
	\includegraphics[width=2\columnwidth]{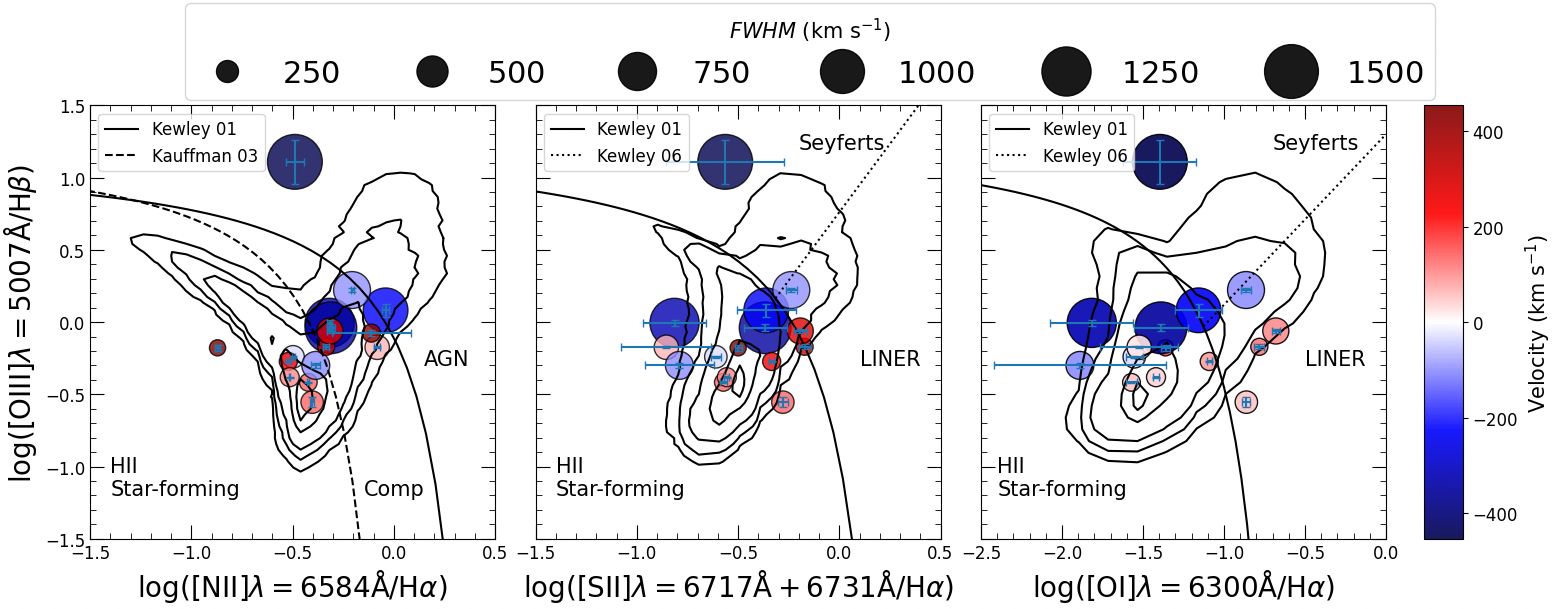}
    \caption{Same as Fig.~\ref{fig:NVSSJ151402+015737_bpt}, for the galaxy
    IRAS 20551-4250, in which case most of the outflowing components are located in the star forming region of the diagram, especially in the [SII]-BPT and [OI]-BPT, which are less affected by metallicity and nitrogen enrichment dependence.}
    \label{fig:IRAS_20551-4250_bpt}
\end{figure*}

In order to study the outflow properties, we analysed the spatially resolved BPT \citep{baldwin1981} and VO \citep{1987ApJS...63..295V} diagrams of the galactic outflows in each galaxy. These diagrams are useful for categorising different regions within a galaxy based on gas excitation and ionisation mechanisms. Specifically, they help identify regions associated with star formation (such as HII regions excited by young, massive hot stars), excitation caused by supermassive accreting black holes (such as Seyfert or Quasar nuclei), and Low Ionisation Nuclear Emission Line Regions (LINERs). The line of demarcation between the various excitation mechanisms, as identified by~\citet[][solid line]{Kewley2001}, \citet[][dashed line]{Kauffmann2003} and~\citet[][dotted line]{kewley2006}, are shown in \Cref{fig:NVSSJ151402+015737_bpt,fig:IRAS_20551-4250_bpt}. The black line contours indicate the distribution of galaxies from the SDSS DR7\footnote{Sloan Digital Sky Survey Data Release 7, available at \url{https://classic.sdss.org/dr7/}}, enclosing 10, 30, 68, 85 and 95~percent of the sample. The flux ratios of the individual gas components extracted at different apertures were calculated and plotted on the diagnostic diagrams shown in \Cref{fig:NVSSJ151402+015737_bpt,fig:IRAS_20551-4250_bpt}. The velocity of each component was calculated with respect to the stellar continuum and denoted by the colour of the marker, ranging from dark red (more redshifted) to dark blue (more blueshifted). The size of the circles corresponds to the {\it FWHM} of the component. The bigger the marker, the larger the width of the emission line component.

Outflows from 4 (F22491-1808, NVSSJ151402+015737, LEDA 166649 and NGC 7130) out of the 7 galaxies analysed are consistent with the typical AGN excitation. \Cref{fig:NVSSJ151402+015737_bpt} shows one of these cases, specifically for galaxy NVSSJ151402+01573, where outflows (darker blue and bigger markers) are in the AGN excitation region (its fitted spectrum is shown in~\Cref{fig:spec_NVSSJ151402+015737,fig:zoom_NVSSJ151402+015737}).

\Cref{fig:IRAS_20551-4250_bpt} instead shows the case
of the galaxy IRAS 20551-4250, for which the majority of markers representing the outflows are located in the star-forming regions. The area between the solid line and dashed line in the [NII]-BPT diagram is often called the "composite region", where SF and AGN are likely to coexist. However, one should note that out of the three types of BPT diagrams, [NII]-BPT diagram is the least reliable one because of its strong dependence on the ionisation parameter~\citep{Strom2017} and, most importantly, on the abundance of nitrogen~\citep{Masters2016}. Nevertheless, most outflow components lying in the SF region of the [SII]-BPT and [OI]-BPT diagram is a strong indicator of star formation in these outflows. We found another two galaxies, IRAS 13120-5453 (\Cref{fig:bpt_IRAS_13120-5453}) and F13229-2934 (\Cref{fig:bpt_F13229-2934}), which have some components of the outflows in the star forming region of the BPT diagrams.


\subsection{{\it In situ} versus external photoionisation}\label{section:photoion}
\begin{figure}
	\includegraphics[width=\columnwidth]{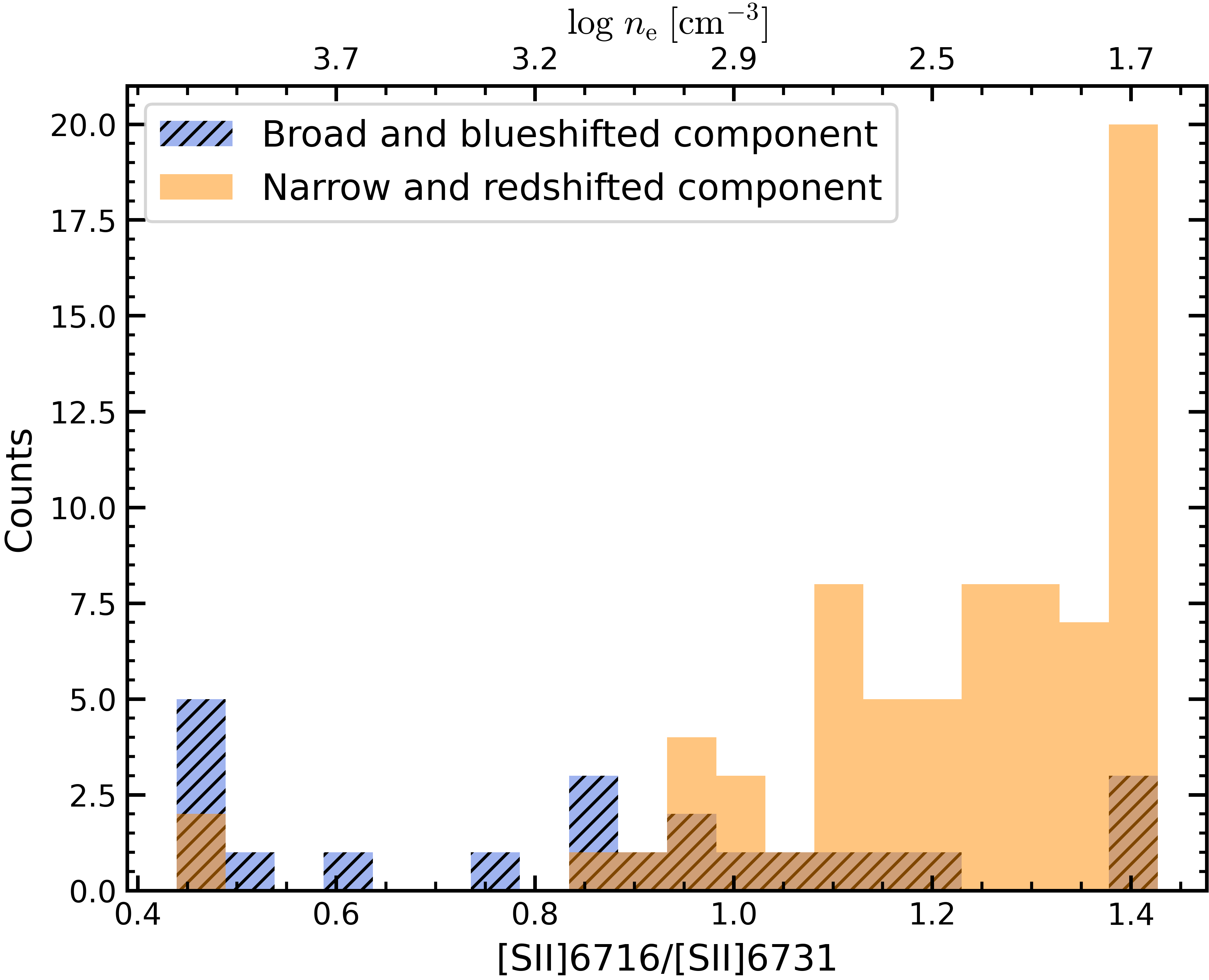}
    \caption{The broad components (blue hatched) tend to have lower [SII] flux ratio, hence higher density than the narrow components (orange). The histogram shows 7 galaxies analysed.}
    \label{fig:s2}
\end{figure}

\begin{figure}
	\includegraphics[width=\columnwidth]{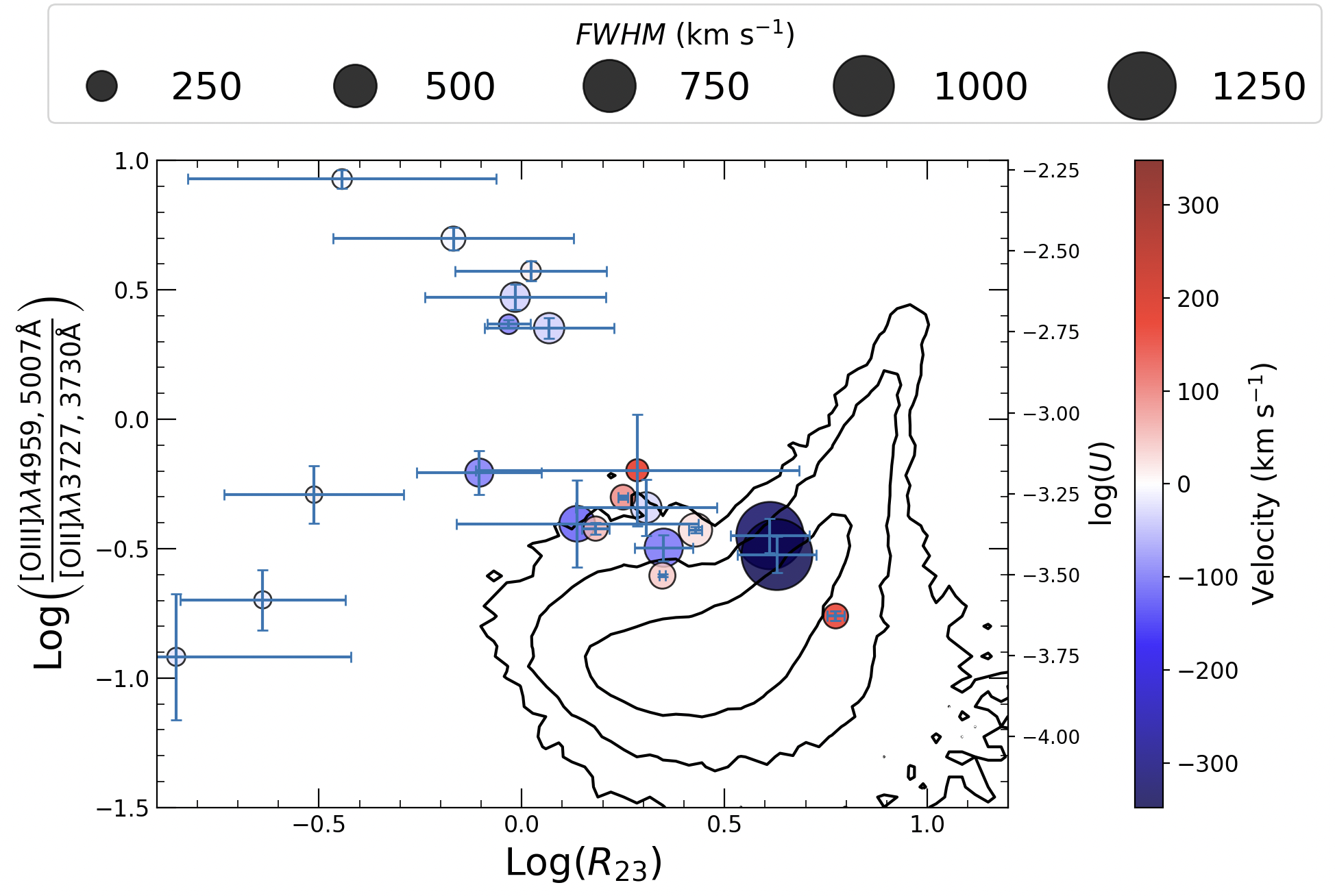}
    \caption{Ionisation parameter inferred from the $F(\mathrm{[OIII]}\lambda\lambda4959,5007\angstrom)$/$F(\mathrm{[OII]}\lambda\lambda3726,3729\angstrom)$ line ratio as a function of the $R_{\mathrm{23}}$ parameter. Circles represent kinematic components in the galaxies where outflows occupy the SF region of the BPT and VO diagrams,  namely IRAS 20551-4250 (robust evidence; see~\Cref{fig:IRAS_20551-4250_whole,fig:IRAS_20551-4250_zoom,fig:IRAS_20551-4250_bpt}), IRAS 13120-5453 (tentative evidence; see~\Cref{fig:spec_IRAS_13120-5453,fig:zoom_IRAS_13120-5453,fig:bpt_IRAS_13120-5453}) and F13229-2934 (tentative evidence; see~\Cref{fig:spec_F13229-2934,fig:zoom_F13229-2934,fig:bpt_F13229-2934}). Outflows (large blue circles) share similar locations with normal star-forming regions (small red circles).
    This result argues strongly in favour of {\it in situ} stellar photoionisation by young stars formed inside the outflows, because in the alternative case where outflows are photoionised by external star formation the blue markers should lie much lower than the red markers. The black line contours indicate the distribution of galaxy population (30, 80 and 95~percent) from the SDSS. Most components lie within the 95~percent population contours. For those who do not, they are situated on the left hand side of the cluster, i.e. have lower log($R_{23}$) values and hence, higher metallicity. These components also have a higher range of uncertainties.}
    \label{fig:ion}
\end{figure}

A possible concern regarding our findings in~\Cref{section:bpt} is that for the outflow components whose BPT diagnostics indicate excitation from young stars, it is plausible that the ionising young stars are not actually situated within the outflow itself, but rather in the galactic disc. In this scenario, the outflowing gas may just be externally illuminated by UV radiation emitted from the disc, instead of hosting star formation within itself. Fortunately, these situations can be differentiated by utilising the ionisation parameter $U$:

\begin{equation}
    U = \frac{Q_{\mathrm{ion}}}{4\pi r^{2} c n_{\mathrm{e}}},
\end{equation}
where $Q_{\mathrm{ion}}/4\pi r^2$ is the ionising photons flux, $n_{\mathrm{e}}$ is the electronic density of the gas and $c$ is the speed of light. \Cref{fig:s2} shows that the gas density, which can be inferred from the [SII] doublet, within the outflow (blue) is comparable to or even higher than that in the galactic disc \citep[orange;][]{2017A&A...606A..96P, mingozzi2019, 2020MNRAS.498.4150D, Fluetsch2019}. The situation of external photoionisation, i.e. $r$ is very large and, in particular much larger than in SF regions, the ionising flux would be remarkably lower compared to that of {\it in situ} photoionisation, i.e. {\it in situ} star formation. Accordingly, taking also into account the higher density, in the scenario of external photoionisation, the ionisation parameter would be orders of magnitude lower than what is typically observed in standard star-forming regions. One can use the ratio
$F(\mathrm{[OIII]}\lambda\lambda4959,5007\angstrom)$/$F(\mathrm{[OII]}\lambda\lambda3726,3729\angstrom)$ as proxy of the ionization parameter \citep{Angeles2000}. \Cref{fig:ion} shows that the ionisation parameter of outflow components (bluer and bigger markers) are indistinguishable from that of normal star-forming regions in the same galaxies (redder and smaller markers) along the y-axis, and higher than typically seen in the majority of star forming galaxies, shown with contours from SDSS DR7. This outcome provides strong supporting evidence for {\it in situ} stellar photoionisation within the outflow.

One should note that the [OIII]/[OII] line ratio has a secondary dependence on gas metallicity, and hence, needs to be monitored via the $R_{\mathrm{23}}$ parameter~\citep{Nagao2006}:
\begin{equation}
    R_{\mathrm{23}} = \frac{F(\mathrm{[OIII]}\lambda\lambda4959,5007\angstrom)+F(\mathrm{[OII]}\lambda\lambda3726,3729\angstrom)}{F(\mathrm{H}\beta \ \lambda4861\angstrom)}.
\end{equation}
The $R_{\mathrm{23}}$ parameter is predominantly sensitive to the oxygen abundance and has a secondary dependence on the ionisation parameter. \Cref{fig:ion} displays both of these quantities by plotting $F(\mathrm{[OIII]}\lambda\lambda4959,5007\angstrom)$/$F(\mathrm{[OII]}\lambda\lambda3726,3729\angstrom)$ ratio against the $R_{\mathrm{23}}$ parameter.

One should note that the [OIII]/[OII] ratio, and the $R_{23}$ parameter are potentially affected by reddening. Unfortunately, we cannot reliably measure dust extinction in the outflow via the $H\alpha$/H$\beta$ Balmer decrement, because the blue and red arm slits probe different regions (because of the different width). However, being off-planar, at least blueshifted outflows (i.e. approaching us) are not expected to be affected by significant reddedning.

\Cref{fig:ion} shows also some points with very low $R_{23}$ parameter, these may potentially be some shocked regions (discussed in the next section), or resulting from the tight [OII] doublet decomposition with other components. However, these relatively narrow components at close to systemic velocities are not relevant for the study presented in this paper.

\begin{figure*}
	\includegraphics[width=2\columnwidth]{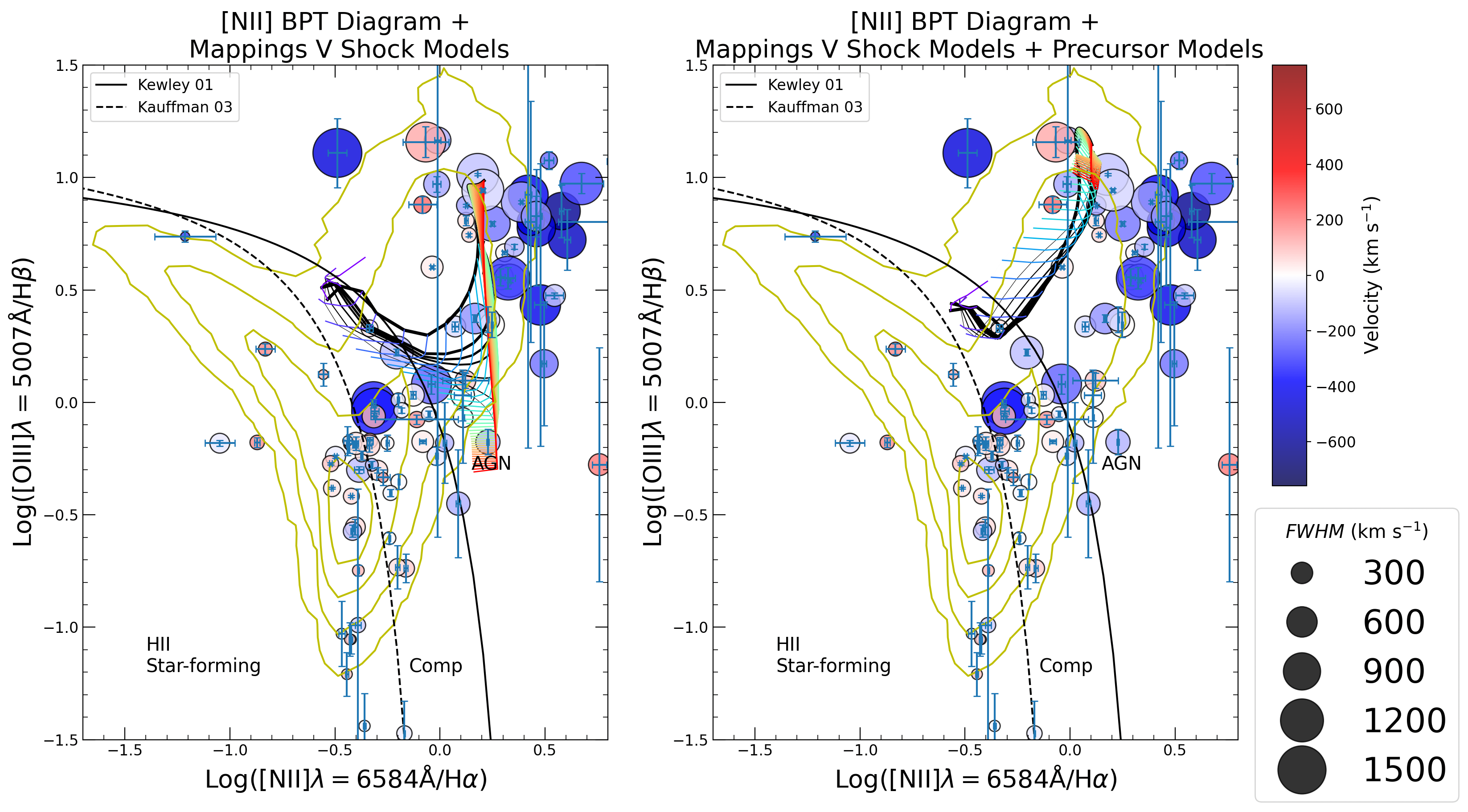}
    \caption{[NII]-BPT diagram displaying the line ratio of gas components for outflows and galactic centre in all 7 of our galaxy samples, as~\Cref{fig:IRAS_20551-4250_bpt} but overlaying grid shock models~\citep{alarie2019} with velocities ranging from 100 km s$^{-1}$
    (purple) to 1000 km s$^{-1}$ (red) and increasing magnetic field parameter from $B/\sqrt{n}=10^{-4} \upmu$ G cm$^{3/2}$ (narrowest line) to $B/\sqrt{n}=10 \upmu$ G cm$^{3/2}$ (thickest line), without (left) and with (right) shock precursor. Galactic outflow components are not affected by the shock models. The yellow line contours enclose the distribution of galaxy population (30\%, 68\%, 95\%, 99\%) from the SDSS.} 
    \label{fig:shock_n2_den1}
\end{figure*}

\begin{figure*}
	\includegraphics[width=2\columnwidth]{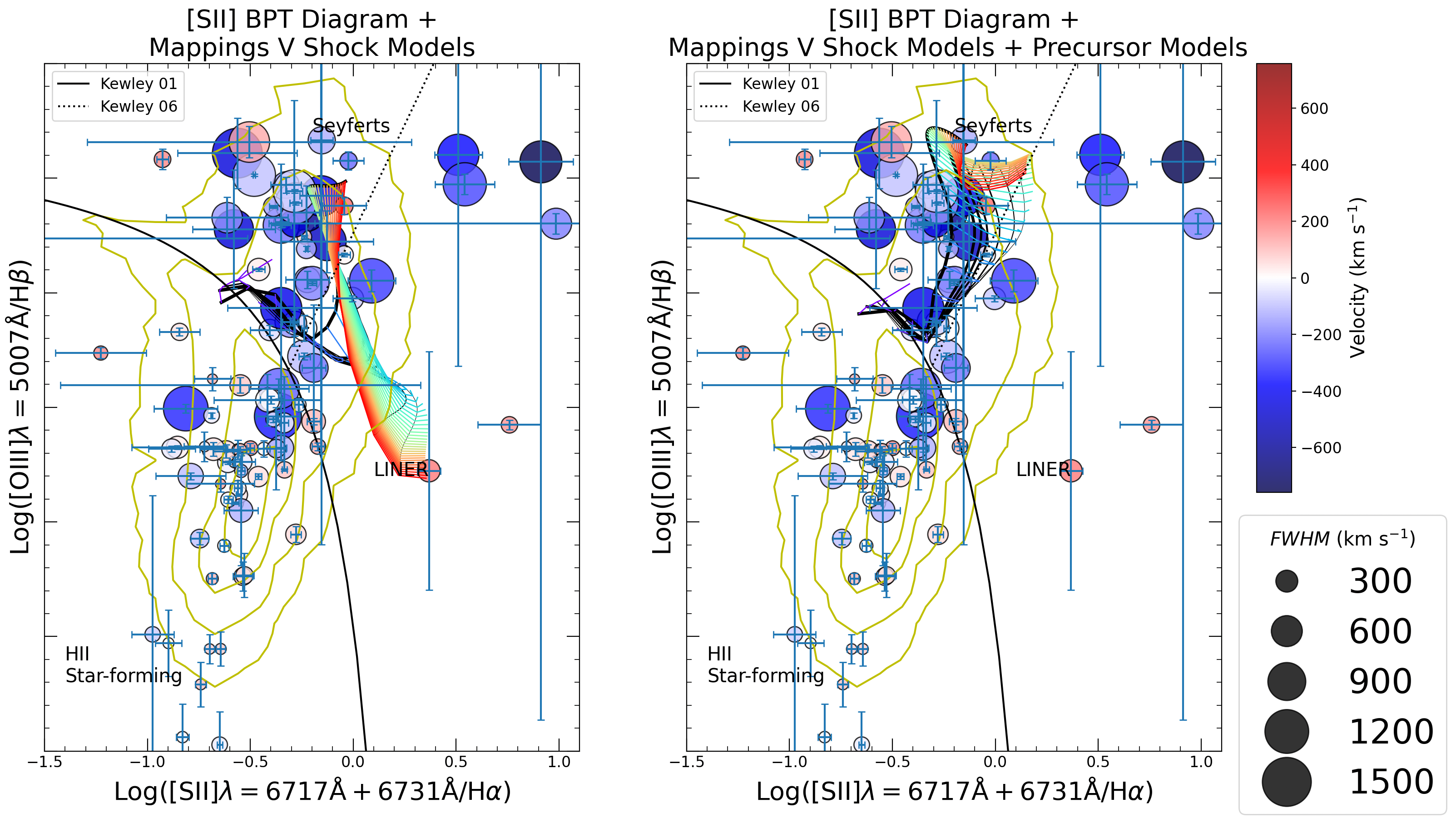}
    \caption{[SII]-BPT diagram displaying the line ratio of gas components for outflows and galactic centre in all 7 of our galaxy samples, as ~\Cref{fig:IRAS_20551-4250_bpt} but overlaying grid shock models~\citep{alarie2019} with velocities ranging from 100 km s$^{-1}$
    (purple) to 1000 km s$^{-1}$ (red) and increasing magnetic field parameter from $B/\sqrt{n}=10^{-4} \upmu$ G cm$^{3/2}$ (narrowest line) to $B/\sqrt{n}=10 \upmu$ G cm$^{3/2}$ (thickest line), without (left) and with (right) shock precursor. Galactic outflow components are not affected by the shock models. The yellow line contours enclose the distribution of galaxy population (30\%, 68\%, 95\%, 99\%) from the SDSS.}  
     \label{fig:shock_s2_den1}
\end{figure*}

\begin{figure*}
	\includegraphics[width=2\columnwidth]{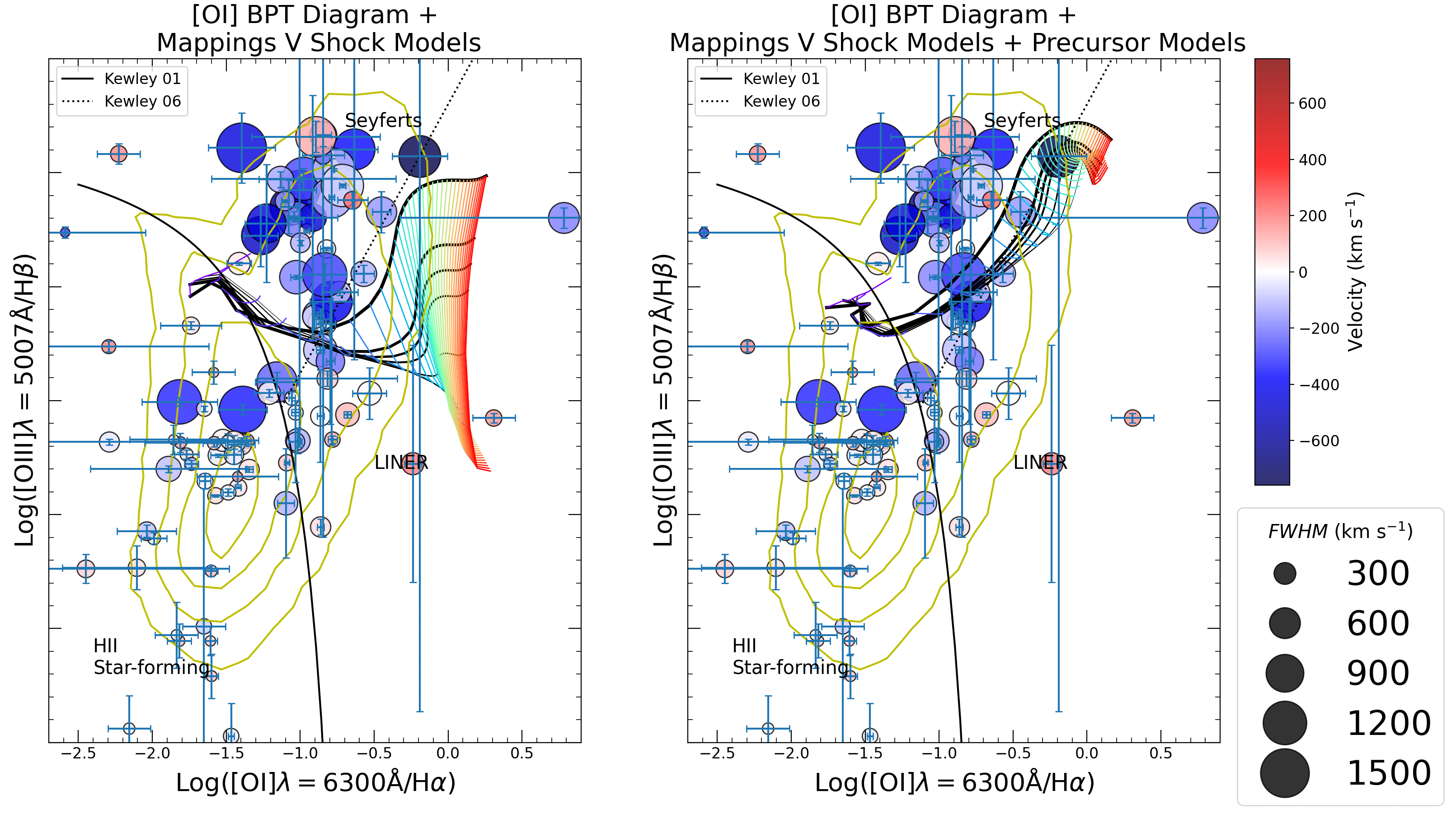}
    \caption{[OI]-BPT diagram displaying the line ratio of gas components for outflows and galactic centre in all 7 of our galaxy samples, as ~\Cref{fig:IRAS_20551-4250_bpt} but overlaying grid shock models~\citep{alarie2019} with velocities ranging from 100 km s$^{-1}$ (purple) to 1000 km s$^{-1}$ (red) and increasing magnetic field parameter from $B/\sqrt{n}=10^{-4} \upmu$ G cm$^{3/2}$ (narrowest line) to $B/\sqrt{n}=10 \upmu$ G cm$^{3/2}$ (thickest line), without (left) and with (right) shock precursor. Galactic outflow components are not affected by the shock models. The yellow line contours enclose the distribution of galaxy population (30\%, 68\%, 95\%, 99\%) from the SDSS.}  
     \label{fig:shock_o1_den1}
\end{figure*}

\subsection{Contribution by shocks}\label{section:shock}
A potential limitation of utilising both the BPT diagrams and the $R_{23}$ vs $O_{32}$ diagram (see~\Cref{fig:ion}) to identify {\it in situ} star formation is that while shocks typically occupy the LINER-like region in these diagrams (e.g.~\citet{Allen2008,Diniz2017}), certain peculiar shocks, particularly those with low velocities, have the potential to produce emission-line ratios resembling star-formation signatures on the BPT diagrams.

To ascertain whether shocks could potentially explain the star-forming-like line ratios on BPT diagrams for the galactic outflows that we have observed, we take a similar approach as~\citet{gallagher2019} by utilising the shock and photoionisation code \textsc{mappings~V}\footnote{\url{https://mappings.anu.edu.au/code/}}~\citep{2018ascl.soft07005S}. We retrieve pre-computed grids from the 
\href{https://sites.google.com/site/mexicanmillionmodels/}{Mexican Million Models Database, 3MdB,}~\citep{alarie2019}. We show extensive grids of these shock models, with and without precursor models, encompassing a broad range of velocities, $v$ = 100 km s$^{-1}$ (purple) to 1000 km s$^{-1}$ (red) and magnetic field parameters, $B/\sqrt{n}=10^{-4} \upmu$ G cm$^{3/2}$ (narrowest line) to $B/\sqrt{n}=10 \upmu$ G cm$^{3/2}$ (thickest line), along with the positions of the line ratios observed in our 7 galaxy samples (see~\Crefrange{fig:shock_n2_den1}{fig:shock_o1_den1} in~\Cref{section:shock}). 

As expected, the shock models are primarily located in the LINER (Low-Ionisation Nuclear Emission-line Region) and AGN-Seyfert regions. While some shock models overlap with some star-forming regions of the diagrams, the majority of the observed line ratios in the star-forming outflows (represented by large blue circles) are inconsistent with the shock models. This further strengthens our assertion that we are indeed witnessing star formation taking place within these outflows.

An additional argument against the shock scenario is that shocks are extremely inefficient in ionizing the gas~\citep{gallagher2019}. Explaining, the observed nebular line luminosities would require that the shocks process huge amount of gas, of the order of several $100~M_\odot/yr$. A more quantitive argument of this issue is given in detail in~\citet{gallagher2019} for galaxies with fainter emission lines in than in our sample, this argument is therefore even stronger for our sample.

\subsection{Star formation rates in outflows}
\label{sec:sfr_results}

Of the three galaxies showing evidence for star formation in their outflows (one with robust evidence, two with tentative evidence), we performed a detailed quantitative analysis on IRAS~20551-4250, which exhibits the strongest evidence. To isolate the highest-velocity gas least likely to be contaminated by disc emission or low-velocity shocks, we adopted a strict outflow criterion ($v < -150$~km/s and FWHM~$>$~200~km/s). Using this definition, we measure a total outflow star formation rate of $\text{SFR}_{\rm out} = 5.24 \pm 0.06~(\text{stat}) \pm 2.62~(\text{sys})$~M$_{\odot}$~yr$^{-1}$. The statistical uncertainty reflects the propagated H$\alpha$ flux measurement errors, while the systematic uncertainty accounts for the calibration of Equation~\ref{eq:sfr}~\citep{kennicutt2012}. This total rate is the sum of two distinct, high-velocity components: one at $v \approx -327$~km/s with FWHM $\approx 1250$~km/s contributing $3.04 \pm 0.04~(\text{stat}) \pm 1.52~(\text{sys})$~M$_{\odot}$~yr$^{-1}$, and a second at $v \approx -348$~km/s with FWHM $\approx 1390$~km/s contributing $2.21 \pm 0.05~(\text{stat}) \pm 1.10~(\text{sys})$~M$_{\odot}$~yr$^{-1}$, where velocities are measured with respect to the stellar continuum.
The total infrared luminosity of IRAS~20551-4250 ($L_{\rm IR} \approx 10^{12.23}~L_{\odot}$) corresponds to a total SFR of $\approx 238$~M$_{\odot}$~yr$^{-1}$~\citep{nardini2010}, assuming a Salpeter initial mass function. However, in ULIRGs like IRAS~20551-4250, a significant fraction of the infrared luminosity originates from dust heated by the AGN rather than star formation.~\citet{nardini2010} determined that the AGN contributes $26 \pm 3$\% to the bolometric luminosity of this system through mid-infrared spectral decomposition. Subtracting this AGN contribution yields a starburst-driven infrared luminosity of $L_{\rm IR,SF} \approx 10^{12.10}~L_{\odot}$ and a corresponding starburst SFR of $\approx 176$~M$_{\odot}$~yr$^{-1}$ (adopting a Chabrier IMF following~\citealp{kennicutt2012}). This AGN-corrected value is consistent with independent estimates based on PAH emission features, which trace star formation with minimal AGN contamination~\citep[e.g., $\approx 200$~M$_{\odot}$~yr$^{-1}$;][]{nardini2010}.

Using the AGN-corrected starburst SFR, the star formation occurring within the high-velocity outflow accounts for approximately $3.0 \pm 1.5$\% of the galaxy's starburst-driven star formation budget (or $2.2 \pm 1.1$\% of the total including the AGN contribution). We stress that the outflow SFR is a robust lower limit, as it is not corrected for internal dust extinction. A contribution of several percent to the total SFR is consistent with findings from other observational studies of star formation in outflows~\citep[e.g.,][]{maiolino2017, gallagher2019} and with predictions from hydrodynamical simulations.

\section{Summary and Conclusions}
Multiple models have put forth the notion that massive galactic outflows could serve as birthplaces for stars, offering intriguing possibilities due to the distinctive kinematic properties of stars formed within these outflows compared to those formed in galactic discs. The observation of substantial quantities of molecular gas, characterised by its high density and clumpiness, within galactic outflows provides compelling evidence supporting the hypothesis that these outflows are indeed conducive to host star formation. 

To investigate how common this new star formation mode is, we used the spectroscopic data obtained from the X-shooter spectrograph to conduct a comprehensive analysis of 12 local (U)LIRG galaxies characterised by the presence of both powerful active galactic nuclei (AGN) and higher levels of star formation. We identify nebular line emission originating from outflows by using \ppxf to perform a kinematic decomposition of the spatially resolved spectra; outflows are identified by their blueshifted velocity and broad line profiles.
We studied the ionisation mechanisms of the outflows using spatially resolved empirical line diagnostics.
Within our subsample of 7 galaxies featuring powerful outflows, we find robust evidence for star formation within the outflow of one galaxy (IRAS 20551-4250), with two additional galaxies showing tentative signs (IRAS 13120-5453 and F13229-2934); the outflow components occupy the star-forming region of the BPT and VO diagrams. Shock-driven photoionisation is rejected based on a comparison of with theoretical models.
We also tested and rejected the hypothesis of external photoionisation by star-forming regions from the underlying galaxy discs, because the ionisation parameter of outflows is indistinguishable from that of normal star-forming regions.
This similarity provides confirmation that the gas within the outflows is not photoionised by the UV emitting from the underlying galactic discs. Instead, it must undergo photoionisation as a result of star formation occurring within the outflows themselves, indicating an {\it in situ} process. Together with previous results
\citep{maiolino2017,gallagher2019,2019MNRAS.486..344R,perna2021,rao2025},
our results suggest that star formation inside galactic outflows may be a relatively common phenomenon.

\section*{Acknowledgements}
The authors would like to express their gratitude to Professor Michele Cappellari, who is heavily involved in the development of the \ppxf package, which is adapted and used to fit our galaxy spectra. The authors are also grateful to Professor Charlie Conroy, who kindly provided the library of stellar templates for stellar continuum fitting. In addition, we are grateful to Alexandre Alarie and Dr Christophe Morisset, who provided the shock models. Authors also acknowledge the contribution of the galaxy population resources from the SDSS project. RM, DDYO and FDE acknowledge support by the Science and Technology Facilities Council (STFC), by the ERC through Advanced Grant 695671 “QUENCH”, and by the UKRI Frontier Research grant RISEandFALL. RM and DDYO also acknowledges funding from a research professorship from the Royal Society.  SA acknowledges grant PID2021-127718NB-I00 funded by the Spanish Ministry of Science and Innovation/State Agency of Research (MICIN/AEI/ 10.13039/501100011033).
AM acknowledges support from PRIN-MUR project "PROME-
TEUS" financed by the European Union - Next Generation EU, Mission 4 Component 1 CUP B53D23004750006. AM acknowledges INAF funding through the "Ricerca Fondamentale 2023" program (mini-grant 1.05.23.04.01). SC acknowledges financial support from the I+D+i project PID2022-140871NB-C21, financed by MICIU/AEI/10.13039/501100011033/ and "FEDER/UE". HR acknowledges support from an Anne McLaren fellowship provided by the University of Nottingham. EB acknowledges support from the Spanish grants PID2022-138621NB-I00 and PID2021-123417OB-I00, funded by MCIN/AEI/10.13039/501100011033/FEDER, EU.
Authors are grateful to Giulia Tozzi who assisted the adaptation of the \ppxf package.

\section*{Data Availability}
The X-shooter spectroscopic data presented in this paper were obtained through the ESO programme 0103.B-0478(A) and 097.B-0918(A) and the associated raw data are
available at the ESO Science Archive Facility: \url{http://archive.eso.org/cms.html}. The raw X-shooter data were reduced and calibrated by the author with the help of the X-shooter data reduction pipeline available at \url{http://www.eso.org/sci/software/pipelines/}. 
The shock models referred to in this paper have been calculated by~\citet{alarie2019} using the shock and photoionisation code mappings V. and is freely accessible from \url{https://mappings.anu.edu.au/code/}.



\bibliographystyle{mnras}
\bibliography{bib} 



\clearpage

\appendix

\section{Spectra fitting and BPT diagnostics for other galaxy samples}\label{bptother}
\begin{figure}
	\centering
	\begin{subfigure}{\textwidth}
        \centering
		\includegraphics[width=0.7\textwidth]{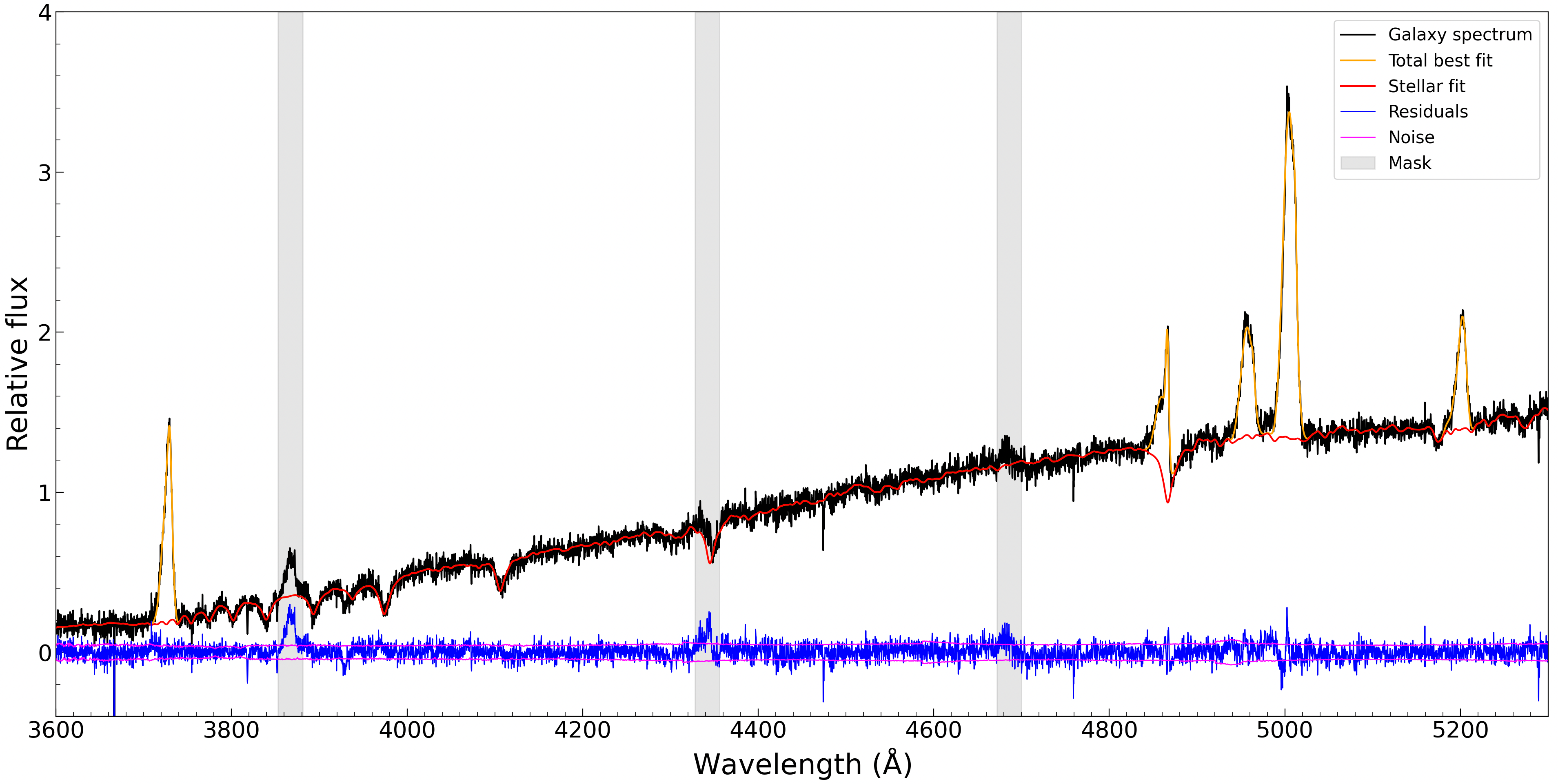}
		\caption{} 
		\label{circle}
	\end{subfigure}
	\begin{subfigure}{\textwidth}
        \centering
		\includegraphics[width=0.72\textwidth]{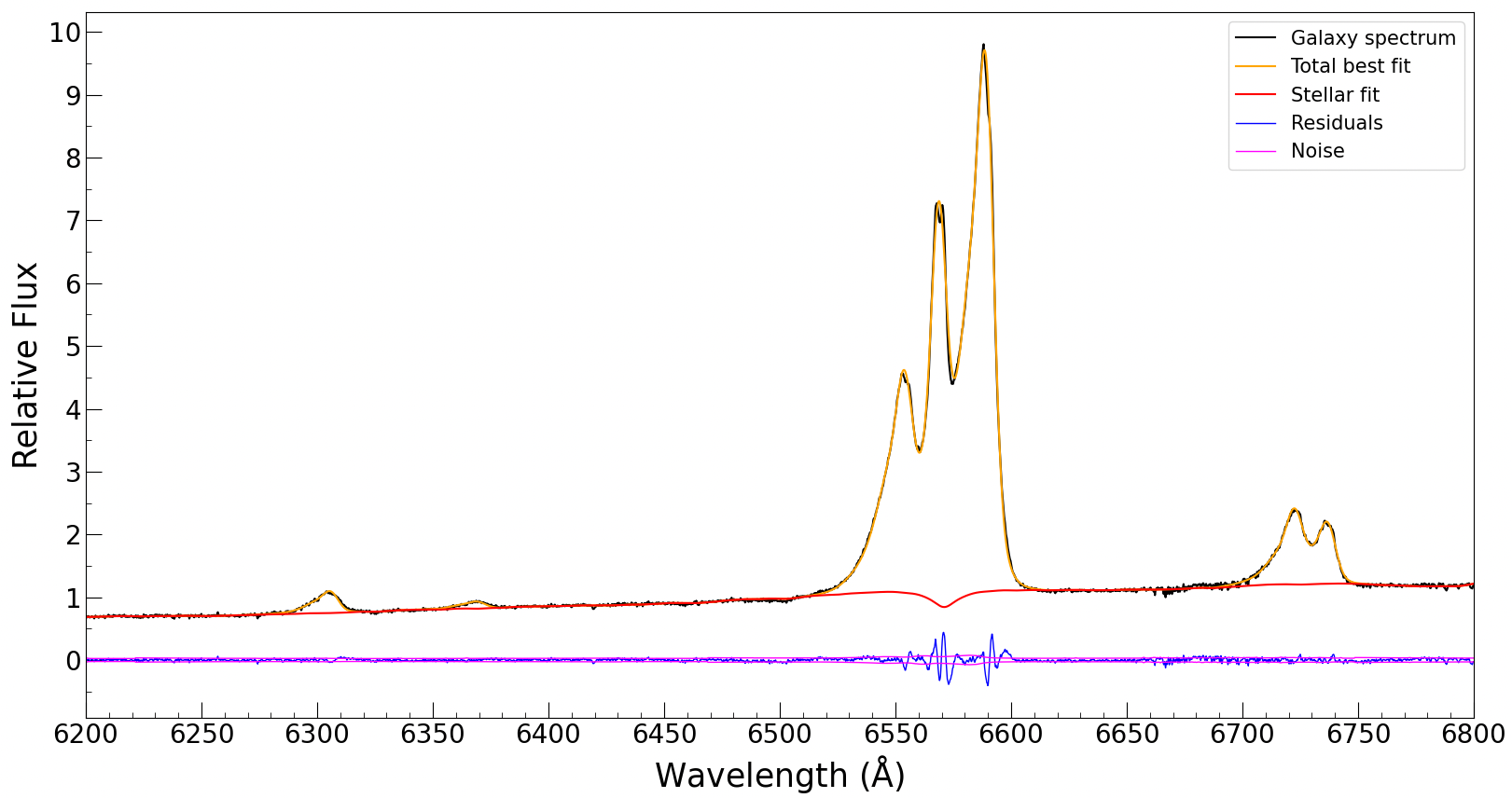}
		\caption{} 
		\label{}
	\end{subfigure}
	\caption{Simultaneously fitting the stellar continuum (red) and emission lines (orange) of the spectrum (black) of IRAS 13120-5453 extracted from the central region in the (a) UVB range and (b) VIS range. Shown below the fit are the residuals (blue) from the \ppxf fitting and the noise spectrum (magenta) from the observation. Emission lines unrelated to the analysis has been masked (grey vertical strips) to avoid interrupting the stellar continuum fitting. Atmospheric absorption has been carefully removed in the VIS range around [SII]6717.} 
    \label{fig:spec_IRAS_13120-5453}
\end{figure}

\begin{figure}
    \includegraphics[width=\columnwidth]{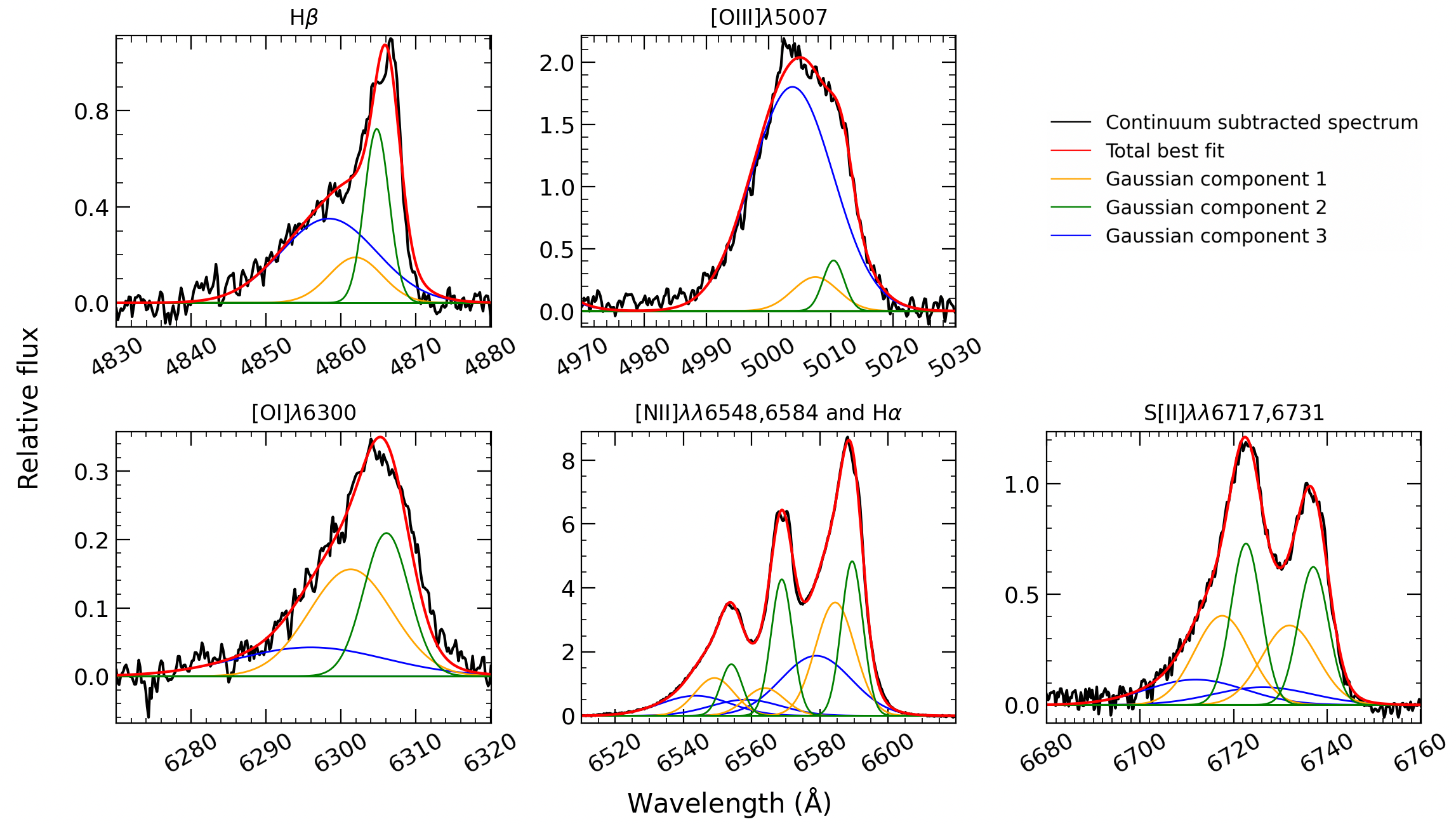}
    \caption{Subsections of continuum-subtracted X-shooter spectra (black) of galaxy IRAS 13120-5453, extracted from the central region around the relevant emission lines for BPT-diagnostics, displaying the decomposition by 3 Gaussian components representing a narrow component (green) associated with the gas in the galactic disk and the two broad components (orange, deep blue) tracing the outflows.}
    \label{fig:zoom_IRAS_13120-5453}
\end{figure}

\begin{figure*}
    \includegraphics[width=\columnwidth]{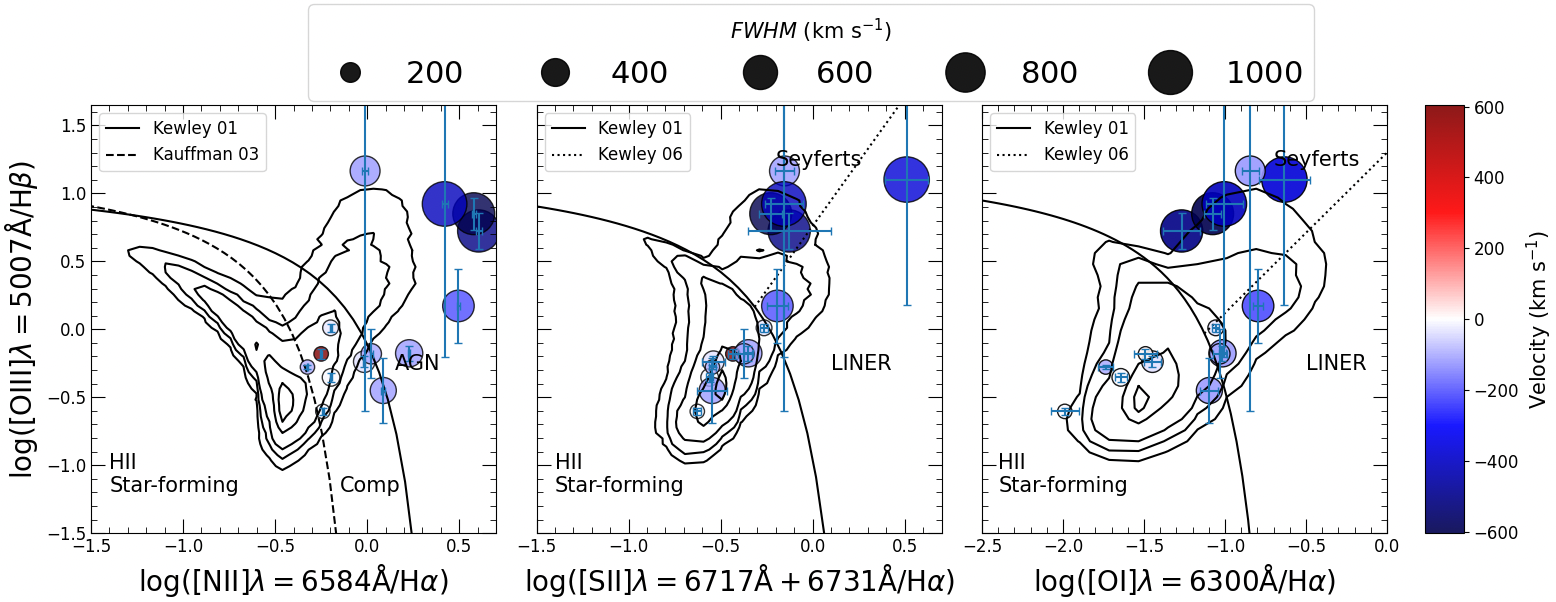}
    \caption{BPT diagnostics diagrams same as~\Cref{fig:NVSSJ151402+015737_bpt}, but for galaxy IRAS 13120-5453. We identify two components, considered as potential outflows due to their very broad widths ($FWHM = 352 \text{\ and } 240 \text{\ km s}^{-1}$) and blueshifted velocities ($v=-114 \text{\ and } -34 \text{\ km s}^{-1}$), which are located in the SF region of the [SII]-BPT diagram and the composite region of [NII]-BPT. The latter one is situated in the SF region of the [OI]-BPT diagram as well. The SF components in these BPT diagrams were tested with the $R_{23}$ vs $O_{23}$ diagram (see~\Cref{fig:ion}) and ruled out the possibility of external photoionisation (see~\Cref{section:photoion}). Contributions from shock have been eliminated as a cause of these SF indicators in~\Cref{section:shock}.}    
    \label{fig:bpt_IRAS_13120-5453}
\end{figure*}

\begin{figure*}
	\centering
	\begin{subfigure}{\textwidth}
        \centering
		\includegraphics[width=0.7\textwidth]{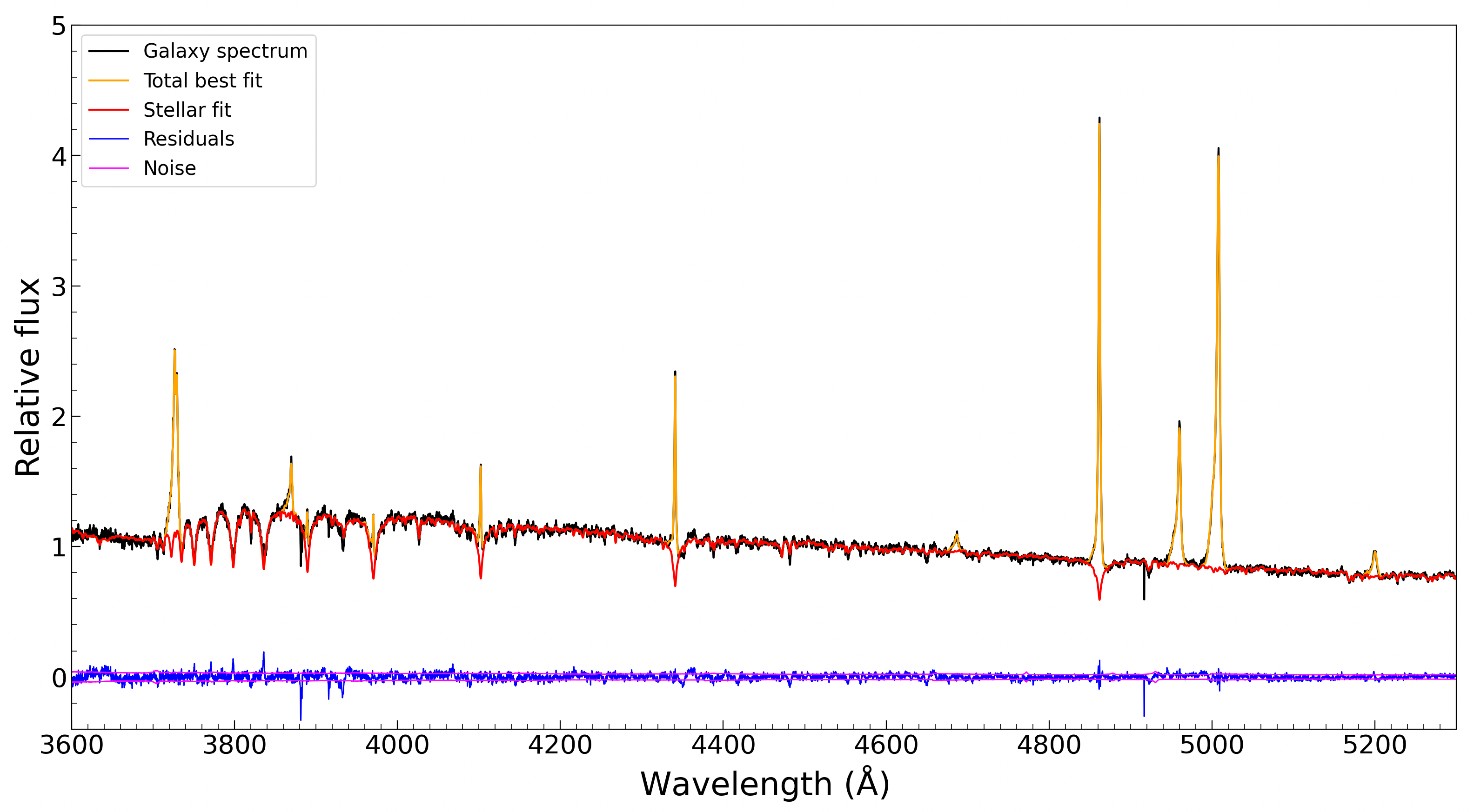}
		\caption{} 
		\label{circle}
	\end{subfigure}
	\begin{subfigure}{\textwidth}
        \centering
		\includegraphics[width=0.72\textwidth]{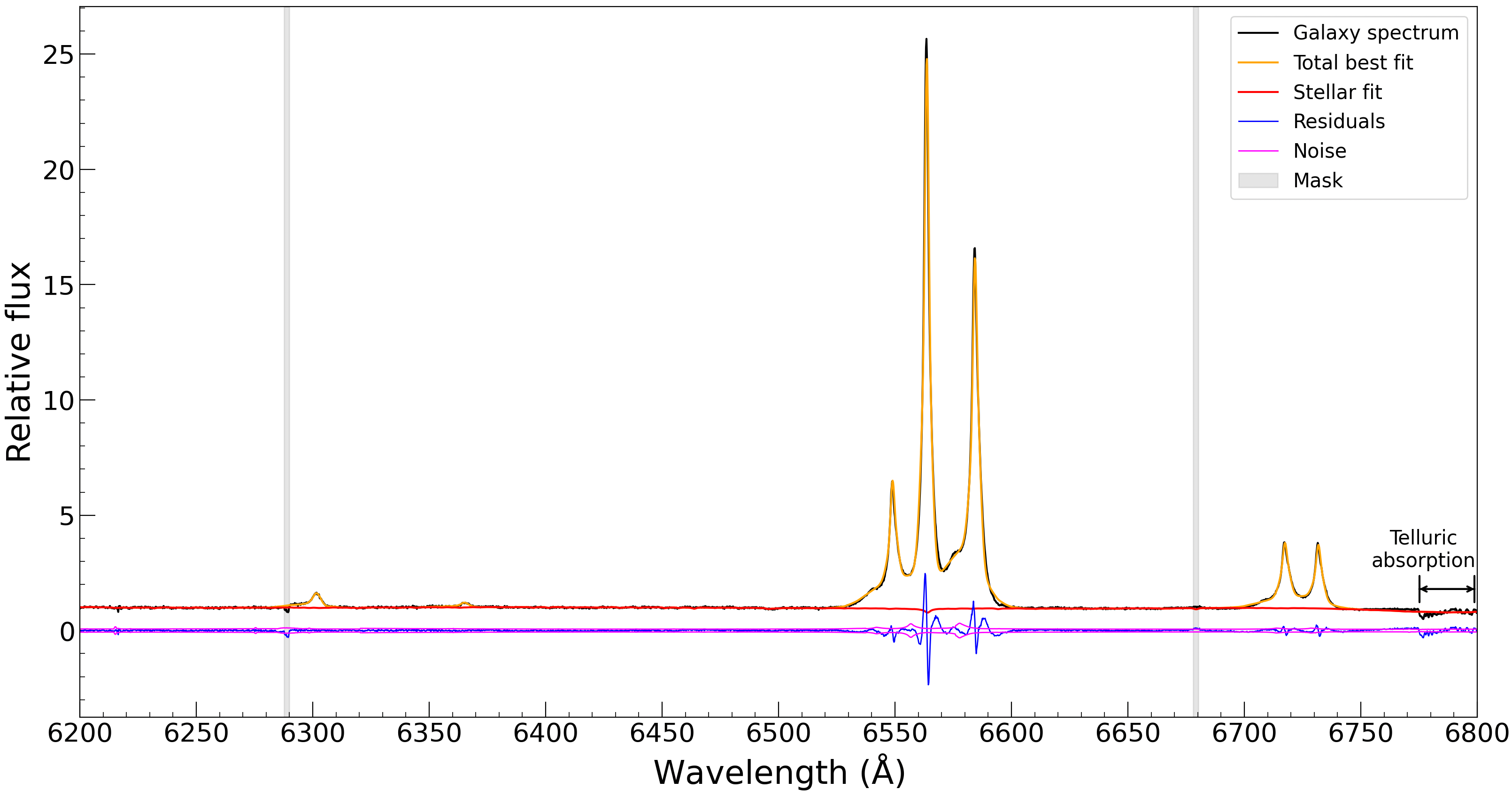}
		\caption{} 
		\label{Mean}
	\end{subfigure}
	\caption{An example of simultaneously fitting the stellar continuum (red) and emission lines (orange) of the spectrum (black) of galaxy F13229-2934 extracted from the central region in the (a) UVB range and (b) VIS range. Shown below the fit are the residuals (blue) from the \ppxf fitting and the noise spectrum (magenta) from the observation. He I line, which is weak and unrelated to the BPT diagnostics, is masked (grey vertical strip) to avoid interrupting the stellar continuum fitting.} 
    \label{fig:spec_F13229-2934}
\end{figure*}

\begin{figure*}
	\includegraphics[width=\columnwidth]{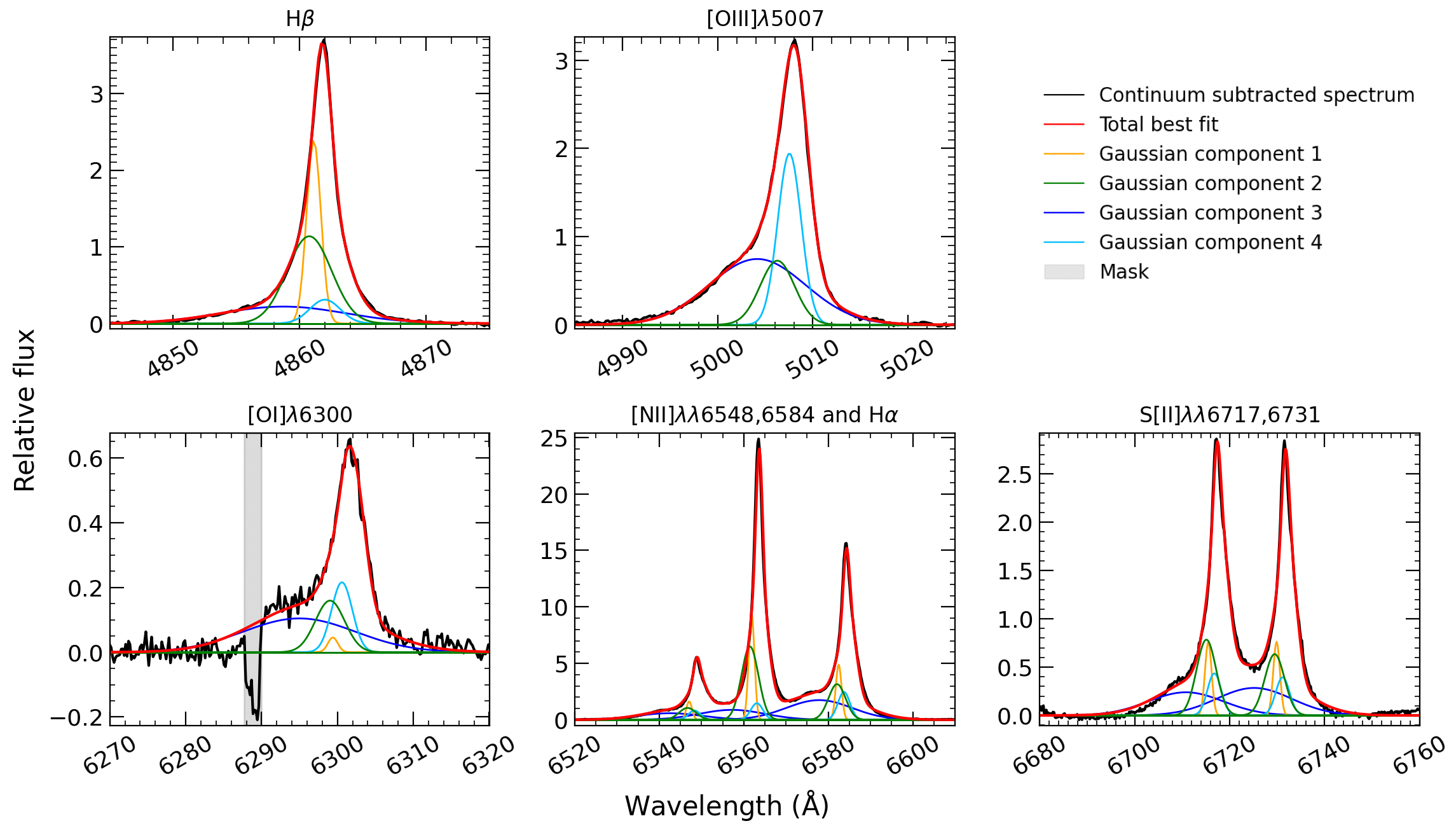}
    \caption{Subsections of continuum-subtracted X-shooter spectra (black) of galaxy F13229-2934, extracted from the central region around the relevant emission lines for BPT-diagnostics, displaying the decomposition by 4 Gaussian components (see legend) representing a narrow component (orange, component 1) associated with the gas in the galactic disk and the broad components (see legend) tracing the outflows. One of the components is not detected by \ppxf for the [OIII]5007 emission line. An upper bound of flux is estimated by 3 times the corresponding flux error estimated by \ppxf.}
    \label{fig:zoom_F13229-2934}
\end{figure*}

\begin{figure*}
	\includegraphics[width=\columnwidth]{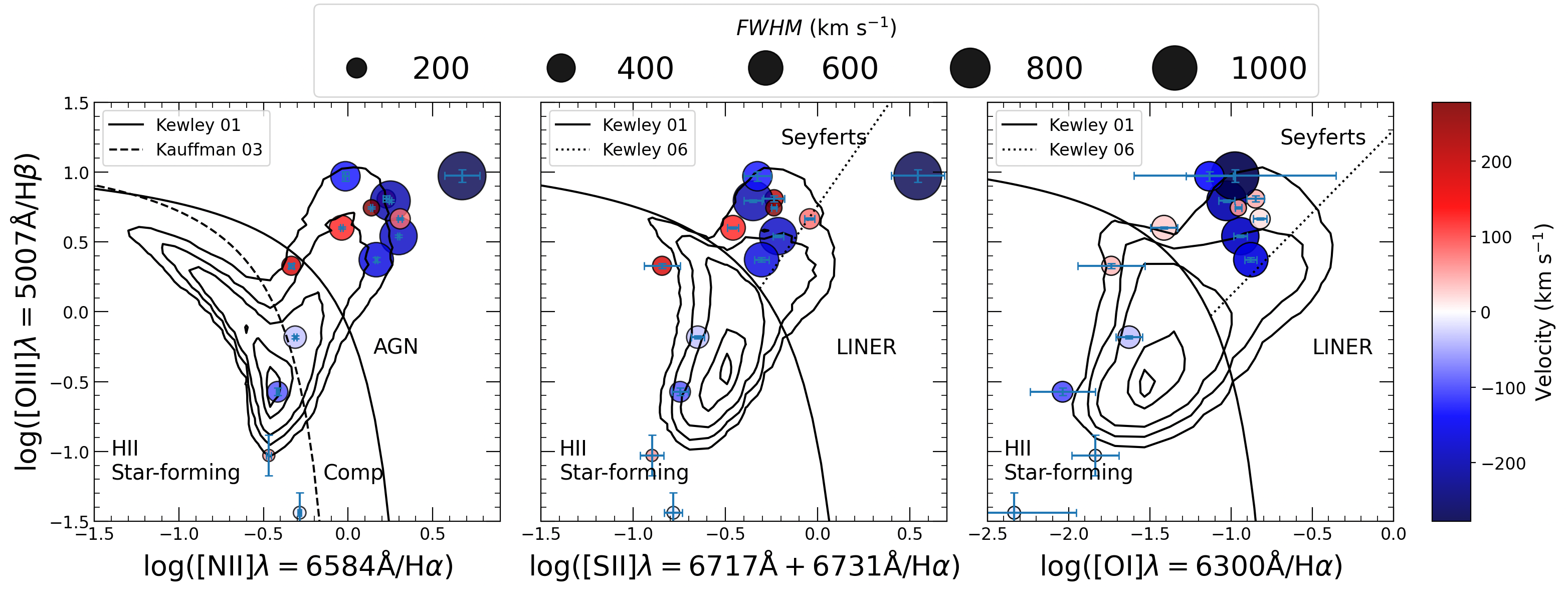}
    \caption{BPT diagnostics diagrams same as~\Cref{fig:NVSSJ151402+015737_bpt}, but for galaxy F13229-2934. We identify two components, considered as potential outflows due to their very broad widths ($FWHM = 255 \text{\ and } 219 \text{\ km s}^{-1}$) and blueshifted velocities ($v=-34 \text{\ and } -97 \text{\ km s}^{-1}$), which are located in the SF region in all BPT diagrams.
    }    
    \label{fig:bpt_F13229-2934}
\end{figure*}

\begin{figure*}
	\centering
	\begin{subfigure}{\textwidth}
        \centering
		\includegraphics[width=0.7\textwidth]{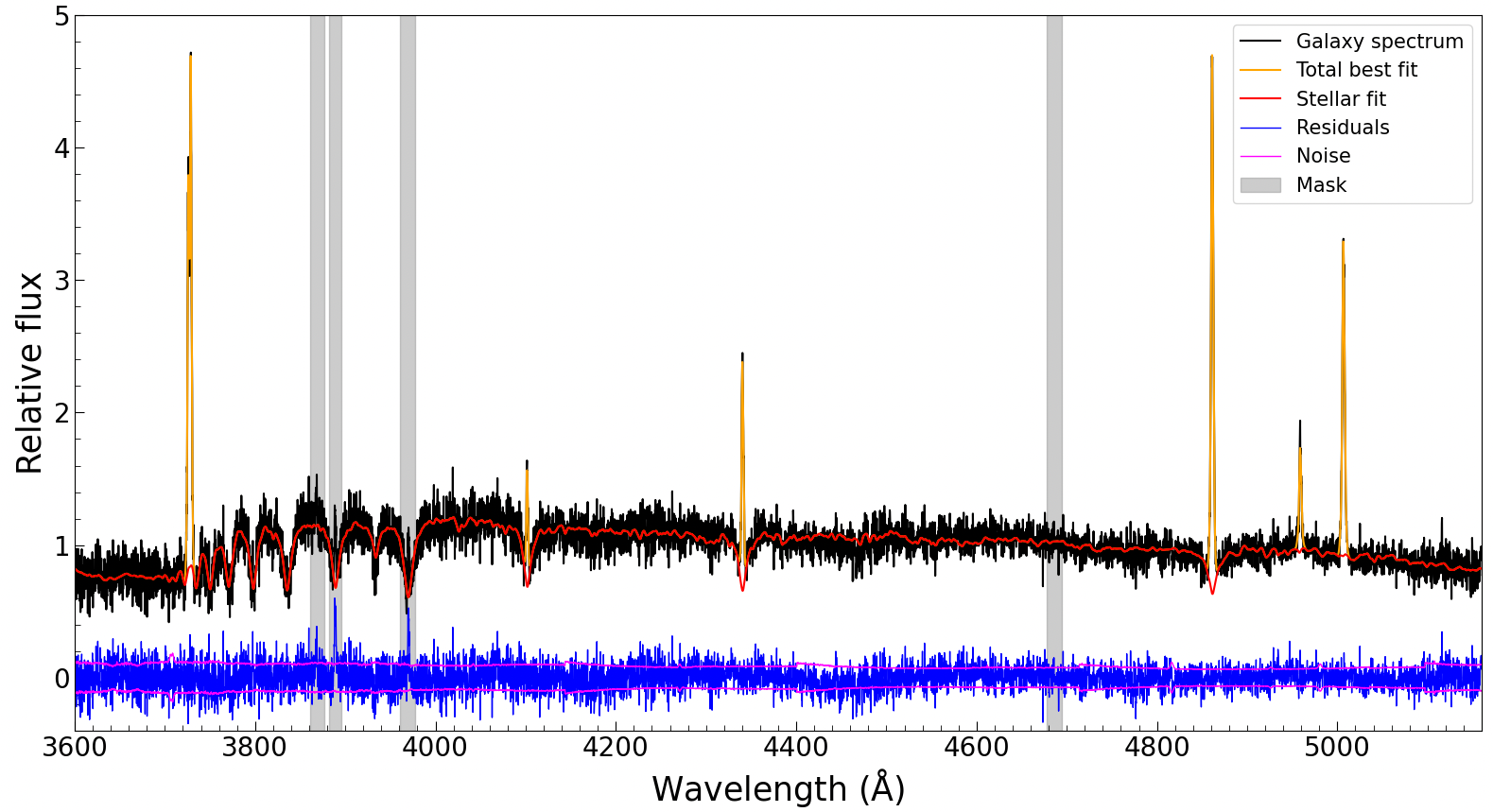}
		\caption{} 
		\label{circle}
	\end{subfigure}
	\begin{subfigure}{\textwidth}
        \centering
		\includegraphics[width=0.72\textwidth]{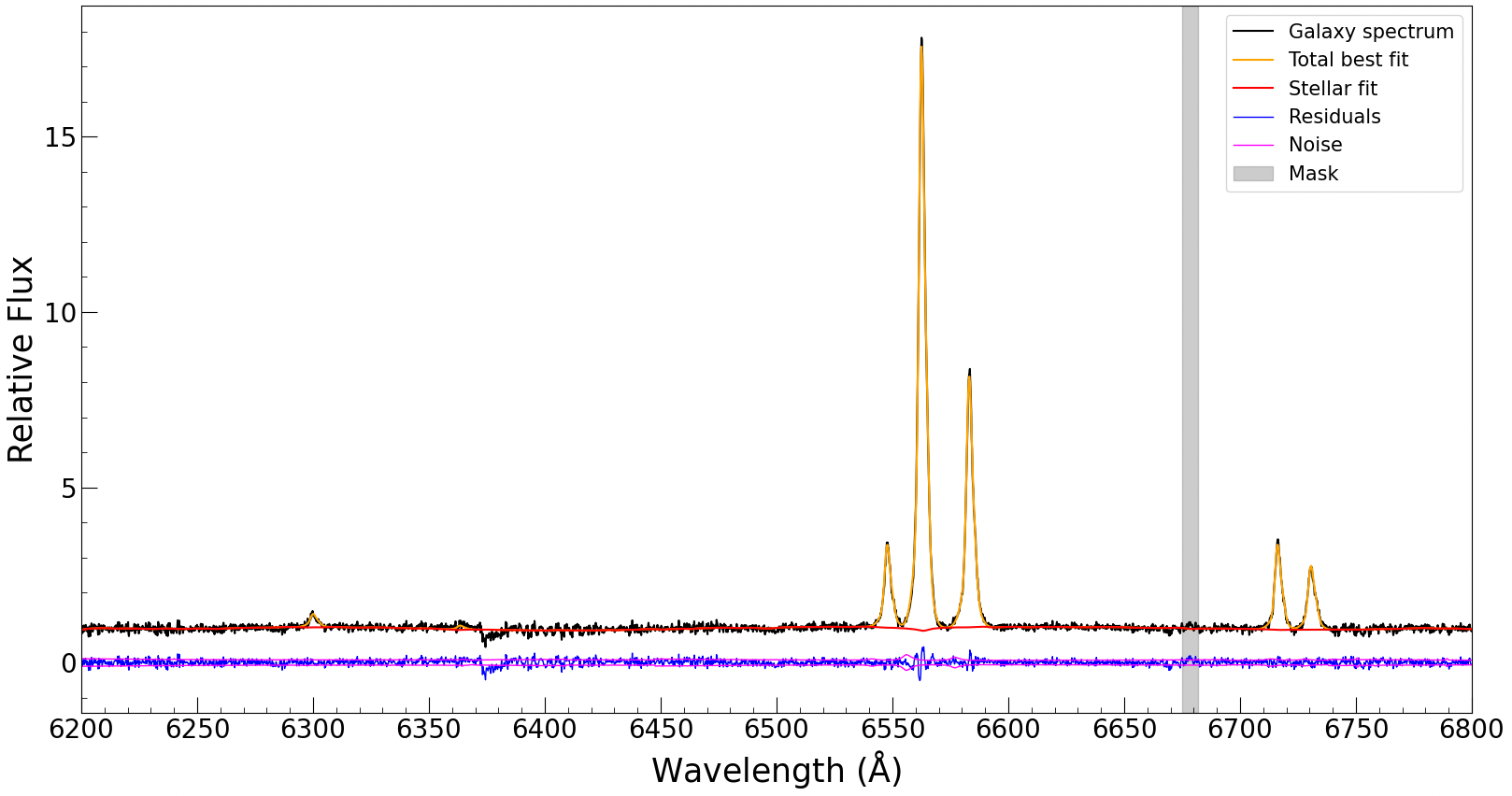}
		\caption{} 
		\label{Mean}
	\end{subfigure}
	\caption{Simultaneously fitting the stellar continuum (red) and emission lines (orange) of the spectrum (black) of galaxy F22491-1808 extracted from the central region in the (a) UVB range and (b) VIS range. Shown below the fit are the residuals (blue) from the \ppxf fitting and the noise spectrum (magenta) from the observation. Emission lines unrelated to the analysis has been masked (grey vertical strips) to avoid interrupting the stellar continuum fitting. Atmospheric absorption has been carefully removed in the VIS range. Due to limitations on visibility during observation of galaxy F22491-1808, the exposure time was only 850 seconds and the seeing was 1.18" (the worst in our data). Fortunately, the S/N ratio is still high enough for spectra fitting and decomposition using 3 Gaussian components (see~\Cref{fig:zoom_F22491-1808}). In the spatially resolved BPT diagnostics, all components are located within the SF region of the BPT diagrams. Most of them are redshifted, while those with negative velocities are too narrow to be considered as outflow components. Based on these information, we suggest that galaxy F22491-1808 may not contain prominent galactic outflows.} 
    \label{fig:spec_F22491-1808}
\end{figure*}

\begin{figure*}
    \includegraphics[width=1\columnwidth]{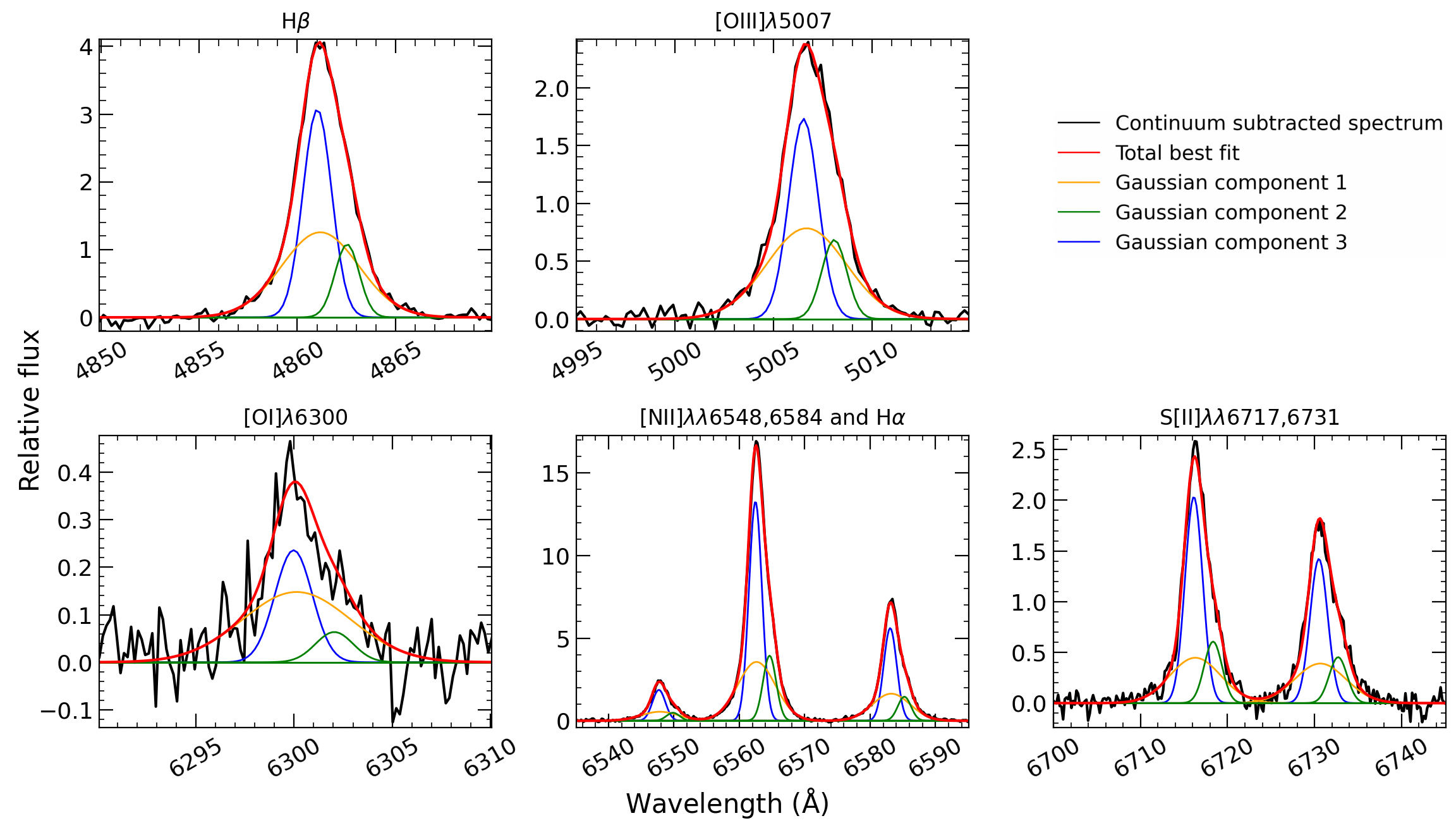}
    \caption{Subsections of continuum-subtracted X-shooter spectra (black) of galaxy F22491-1808, extracted from the central region around the relevant emission lines for BPT-diagnostics, displaying the decomposition of each emission line by 3 Gaussian components (see legend).}
    \label{fig:zoom_F22491-1808}
\end{figure*}

\begin{figure*}
	\includegraphics[width=\columnwidth]{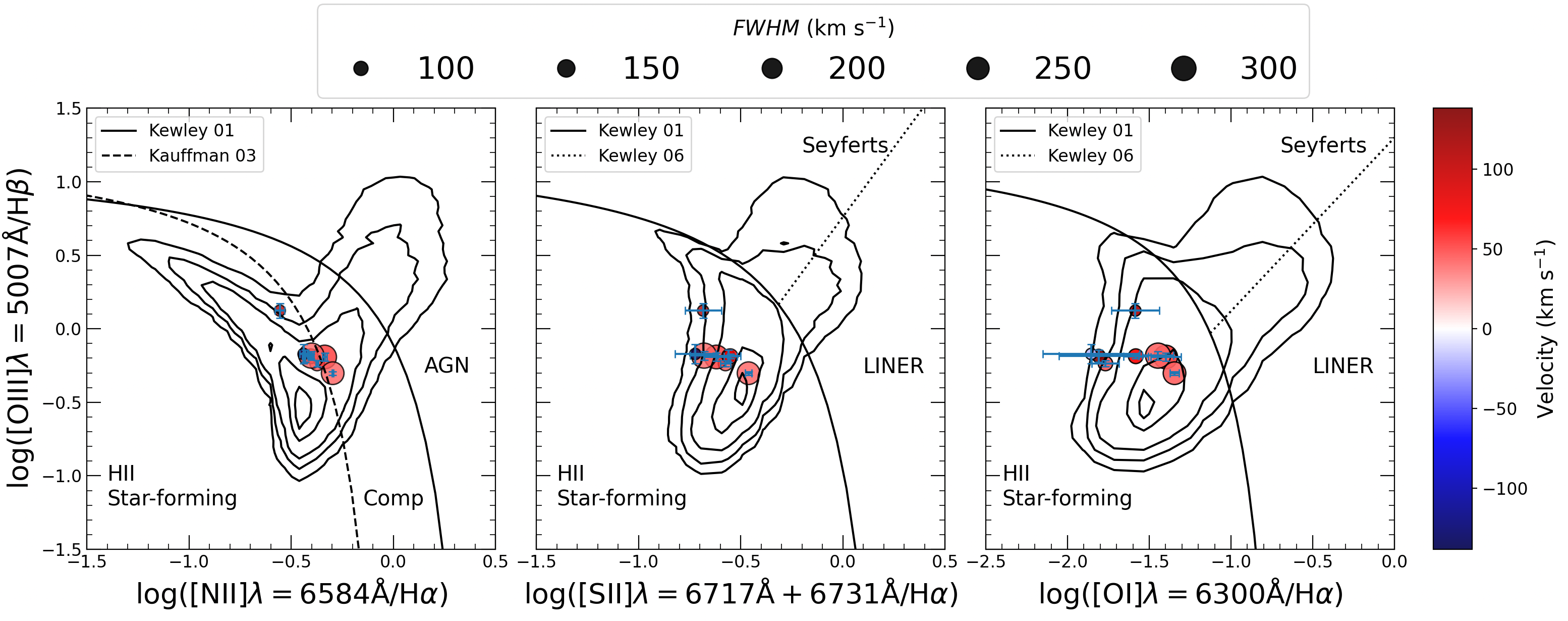}
    \caption{BPT diagnostics diagrams same as~\Cref{fig:NVSSJ151402+015737_bpt}, but for galaxy F22491-1808. All components are located within the SF region of the BPT diagrams. Most of them are redshifted, while those with negative velocities are too narrow to be considered as outflow components. Therefore, galaxy F22491-1808 may not contain prominent galactic outflows.}    
    \label{fig:bpt_F22491-1808}
\end{figure*}

\begin{figure*}
	\centering
	\begin{subfigure}{\textwidth}
        \centering
		\includegraphics[width=0.7\textwidth]{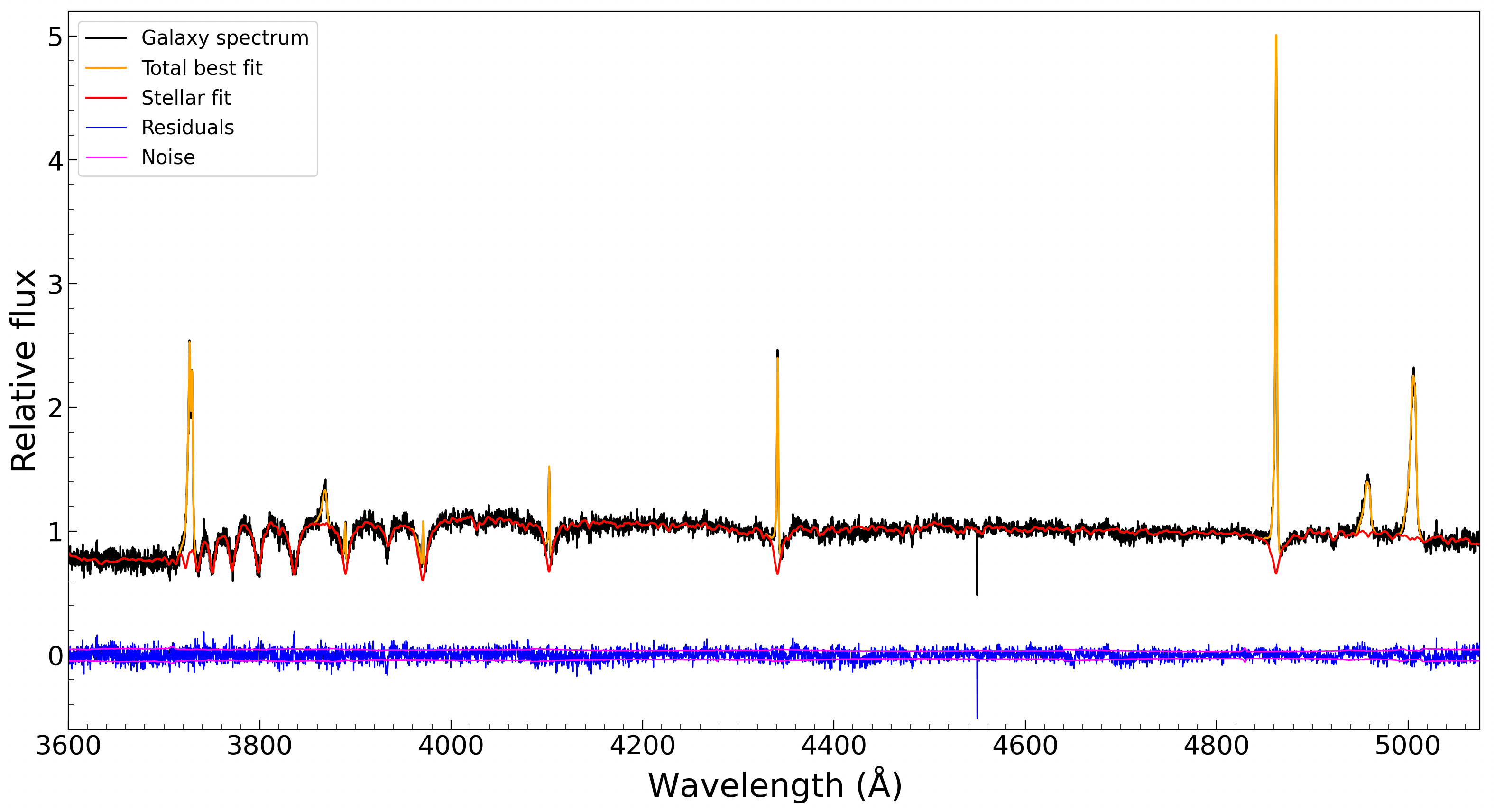}
		\caption{} 
		\label{circle}
	\end{subfigure}
	\begin{subfigure}{\textwidth}
        \centering
		\includegraphics[width=0.72\textwidth]{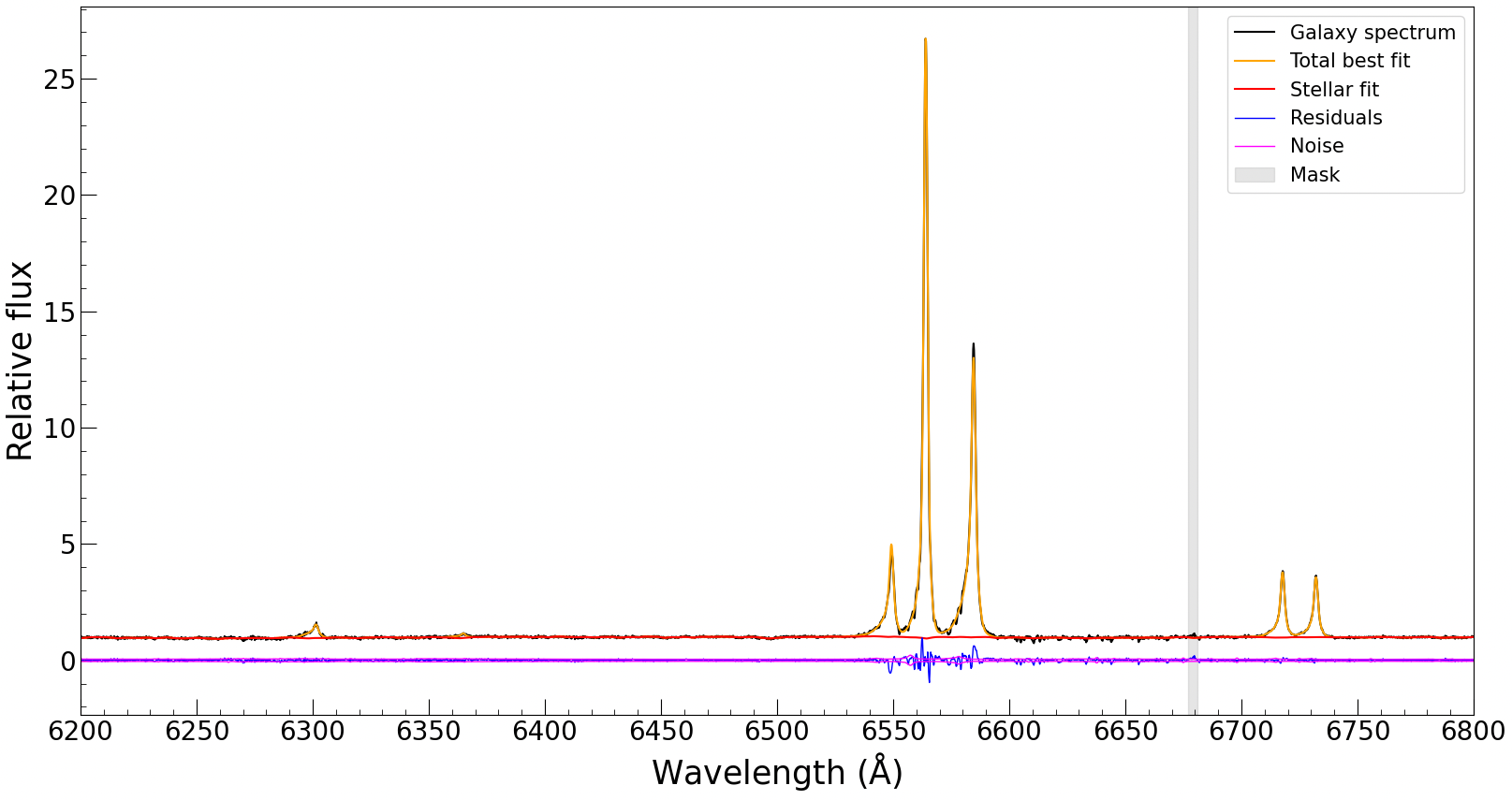}
		\caption{} 
		\label{Mean}
	\end{subfigure}
	\caption{An example of simultaneously fitting the stellar continuum (red) and emission lines (orange) of the spectrum (black) of galaxy NVSSJ151402+015737 extracted from the central region in the (a) UVB range and (b) VIS range. Shown below the fit are the residuals (blue) from the \ppxf fitting and the noise spectrum (magenta) from the observation. Atmospheric absorption has been corrected around [OI]6300 and He I line is masked during PPXF fitting.} 
    \label{fig:spec_NVSSJ151402+015737}
\end{figure*}

\begin{figure*}
    \includegraphics[width=\columnwidth]{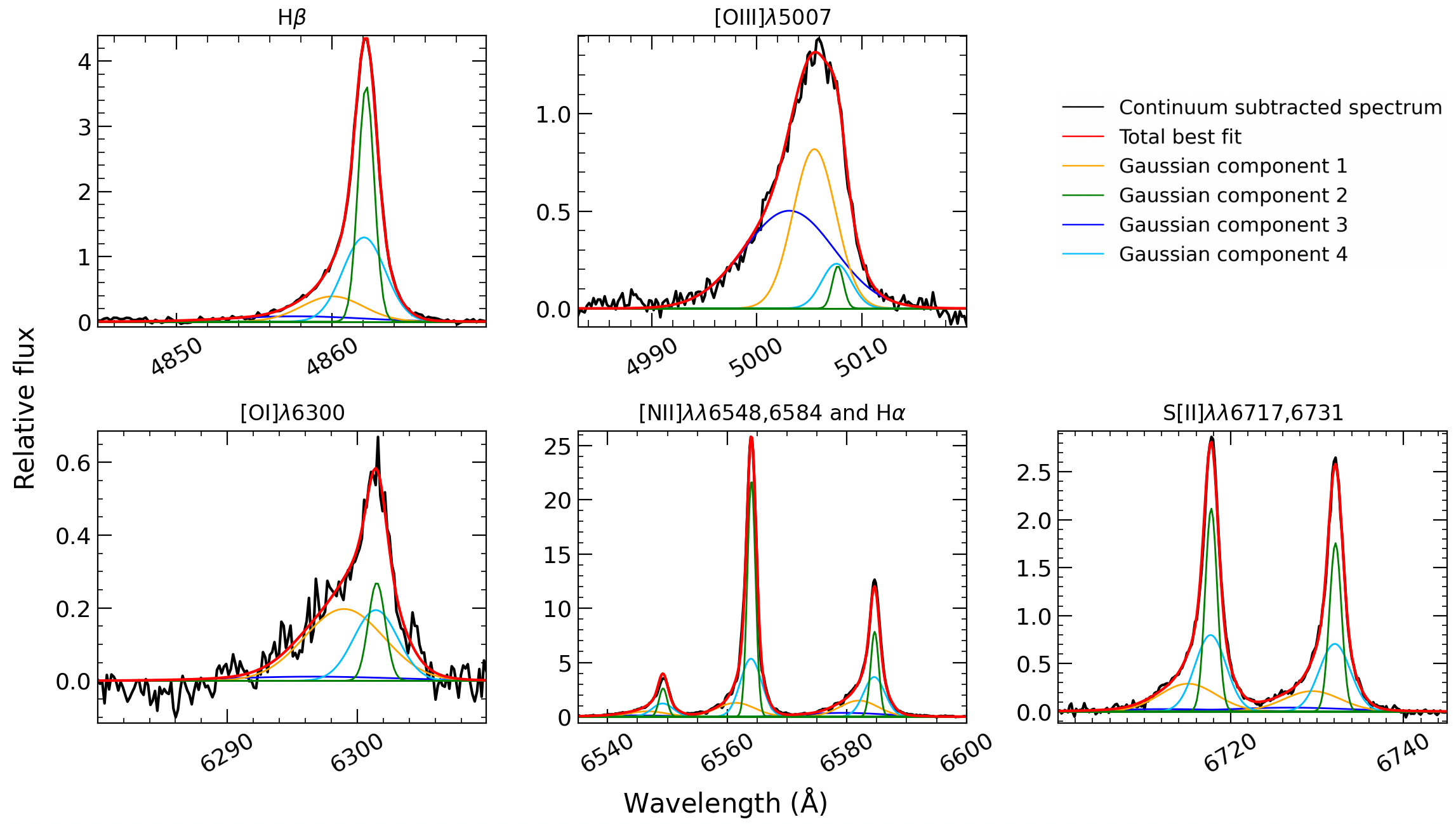}
    \caption{Subsections of continuum-subtracted X-shooter spectra (black) of galaxy NVSSJ151402+015737, extracted from the central region around the relevant emission lines for BPT-diagnostics, displaying the decomposition by 4 Gaussian components representing a narrow component associated with the gas in the galactic disk (component 1) and 3 broad components (component 2 to 4) tracing the outflows.}
    \label{fig:zoom_NVSSJ151402+015737}
\end{figure*}

\begin{figure*}
	\centering
	\begin{subfigure}{\textwidth}
        \centering
		\includegraphics[width=0.7\textwidth]{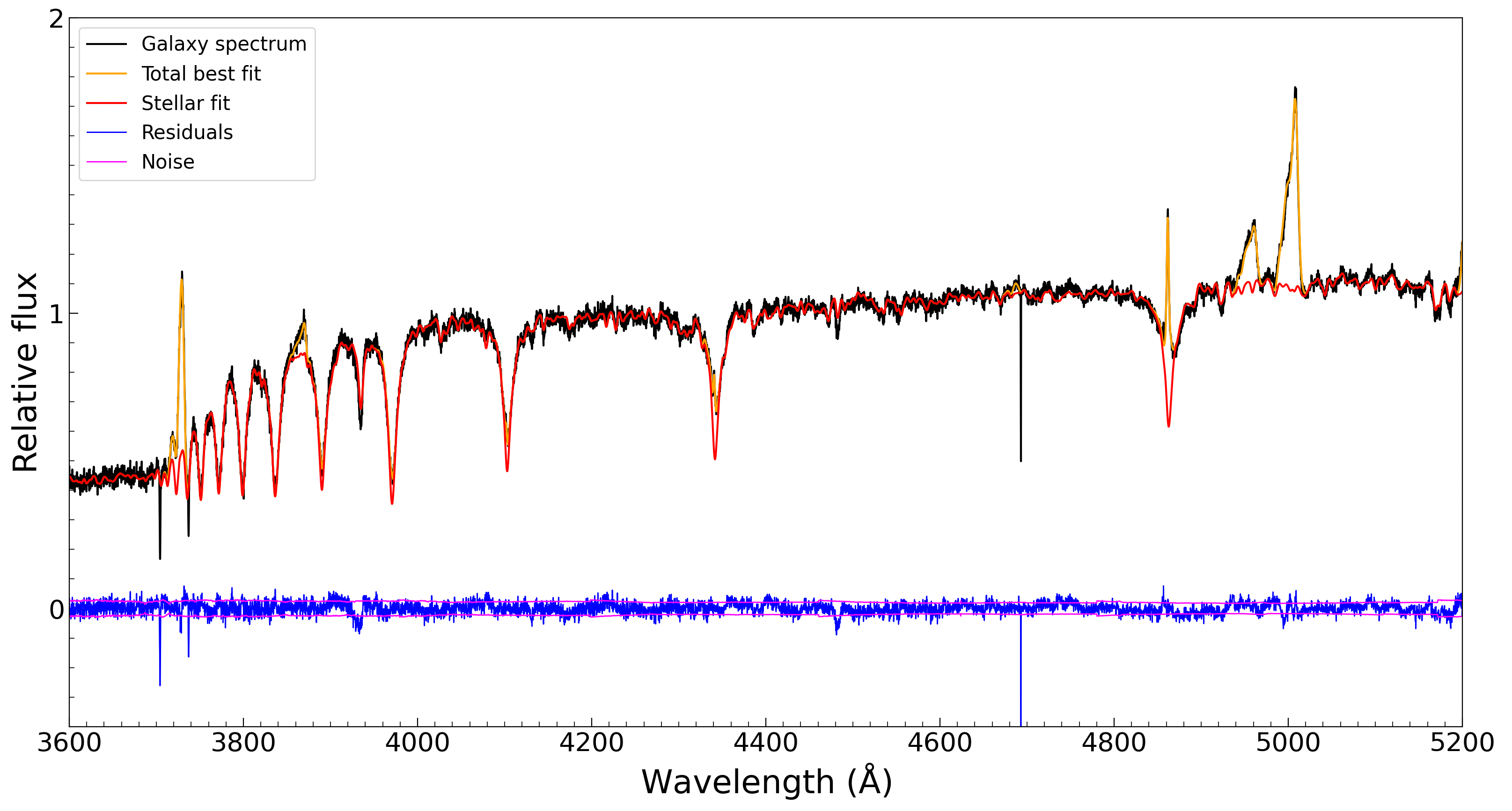}
		\caption{} 
		\label{circle}
	\end{subfigure}
	\begin{subfigure}{\textwidth}
        \centering
		\includegraphics[width=0.72\textwidth]{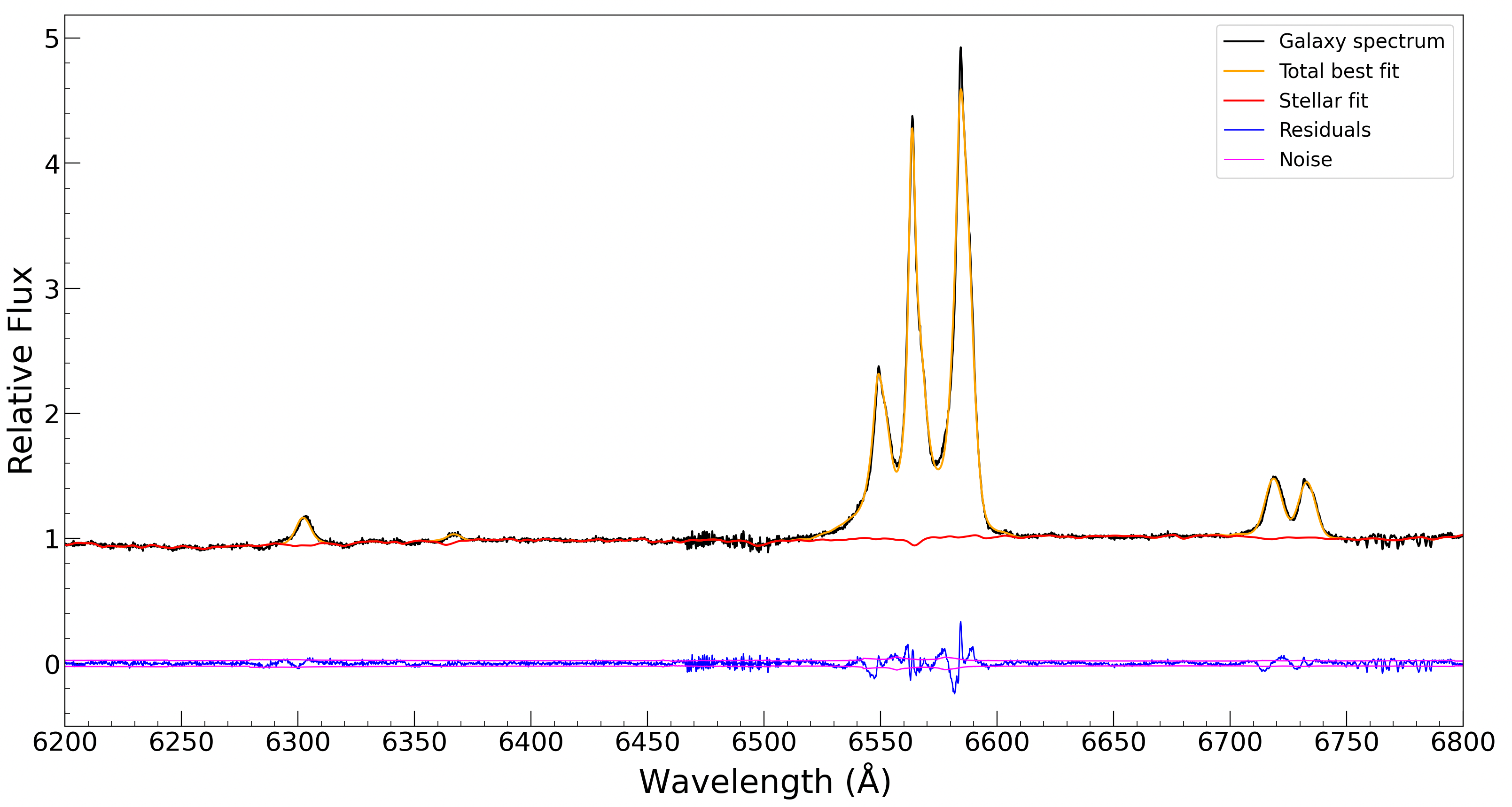}
		\caption{} 
		\label{Mean}
	\end{subfigure}
	\caption{An example of simultaneously fitting the stellar continuum (red) and emission lines (orange) of the spectrum (black) of galaxy LEDA166649 extracted from the central region in the (a) UVB range and (b) VIS range. Shown below the fit are the residuals (blue) from the \ppxf fitting and the noise spectrum (magenta) from the observation. Telluric correction has been performed between $\lambda=6460$ to 6520 $\angstrom$.} 
    \label{fig:spec_LEDA166649}
\end{figure*}

\begin{figure*}
    \includegraphics[width=\columnwidth]{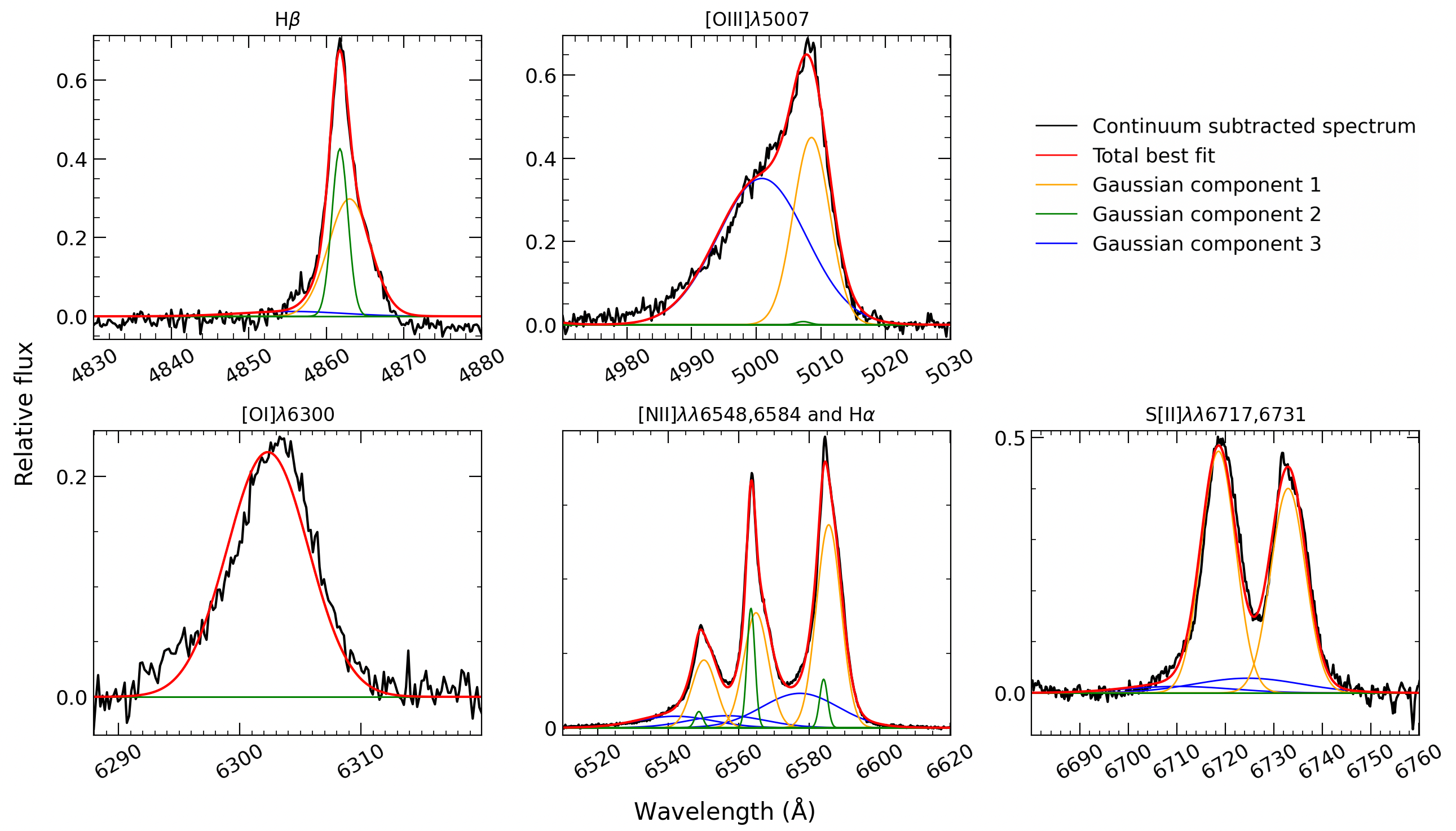}
    \caption{Subsections of continuum-subtracted X-shooter spectra (black) of galaxy LEDA166649, extracted from the central region around the relevant emission lines for BPT-diagnostics, displaying the decomposition by 3 Gaussian components representing the narrow component (green, component 2) associated with the gas in the galactic disk and the broad components (component 1 and 3) tracing the outflows. For the [OI]6300 line, only gaussian component 1 (orange) is detected by PPXF. For the [SII] doublet, gaussian component 2 (green) is not detected. An upper bound of flux for the weak, undetectable component is estimated by 3 times the corresponding flux error determined by \ppxf.}
    \label{fig:zoom_LEDA166649}
\end{figure*}

\begin{figure*}
    \includegraphics[width=\columnwidth]{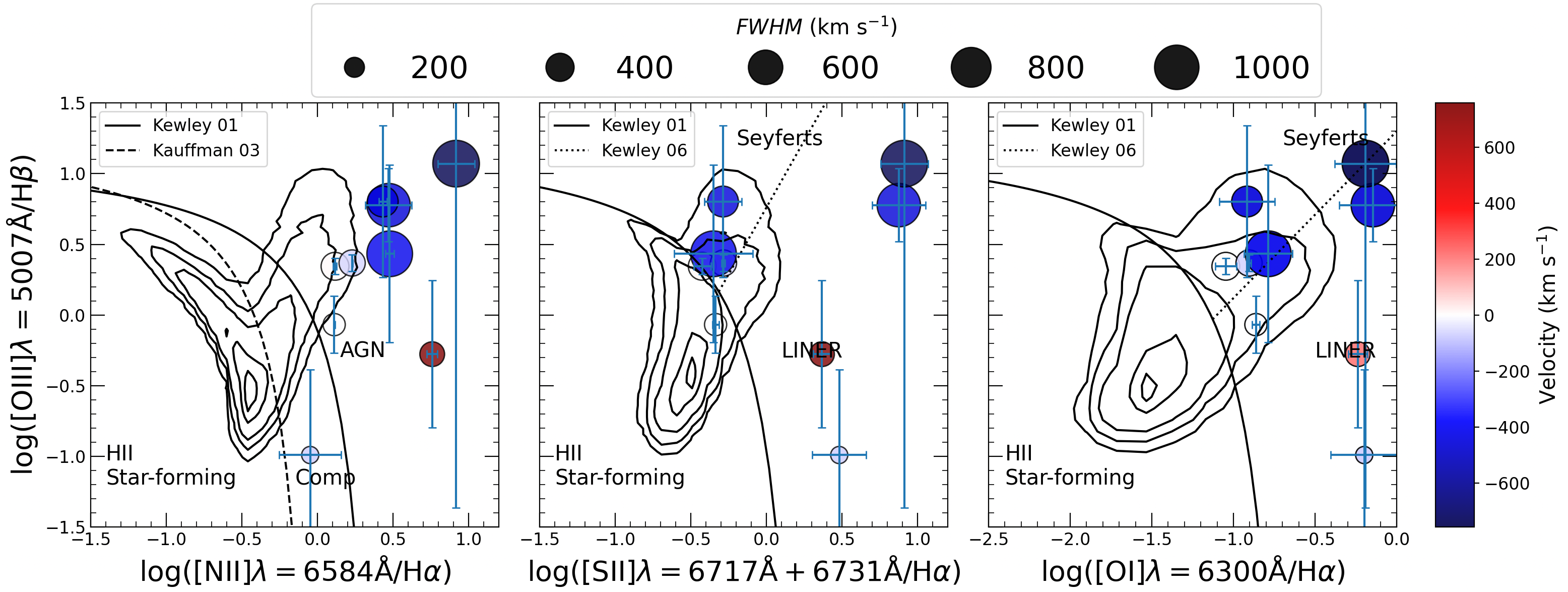}
    \caption{BPT diagnostics diagrams same as~\Cref{fig:NVSSJ151402+015737_bpt}, but for galaxy LEDA166649. None of the broad and blueshifted gas components are located in the SF region. Therefore, galaxy LEDA166649 is highly likely to not host SF in its galactic outflows.}    
    \label{fig:bpt_LEDA166649}
\end{figure*}

\begin{figure*}
	\centering
	\begin{subfigure}{\textwidth}
        \centering
		\includegraphics[width=0.7\textwidth]{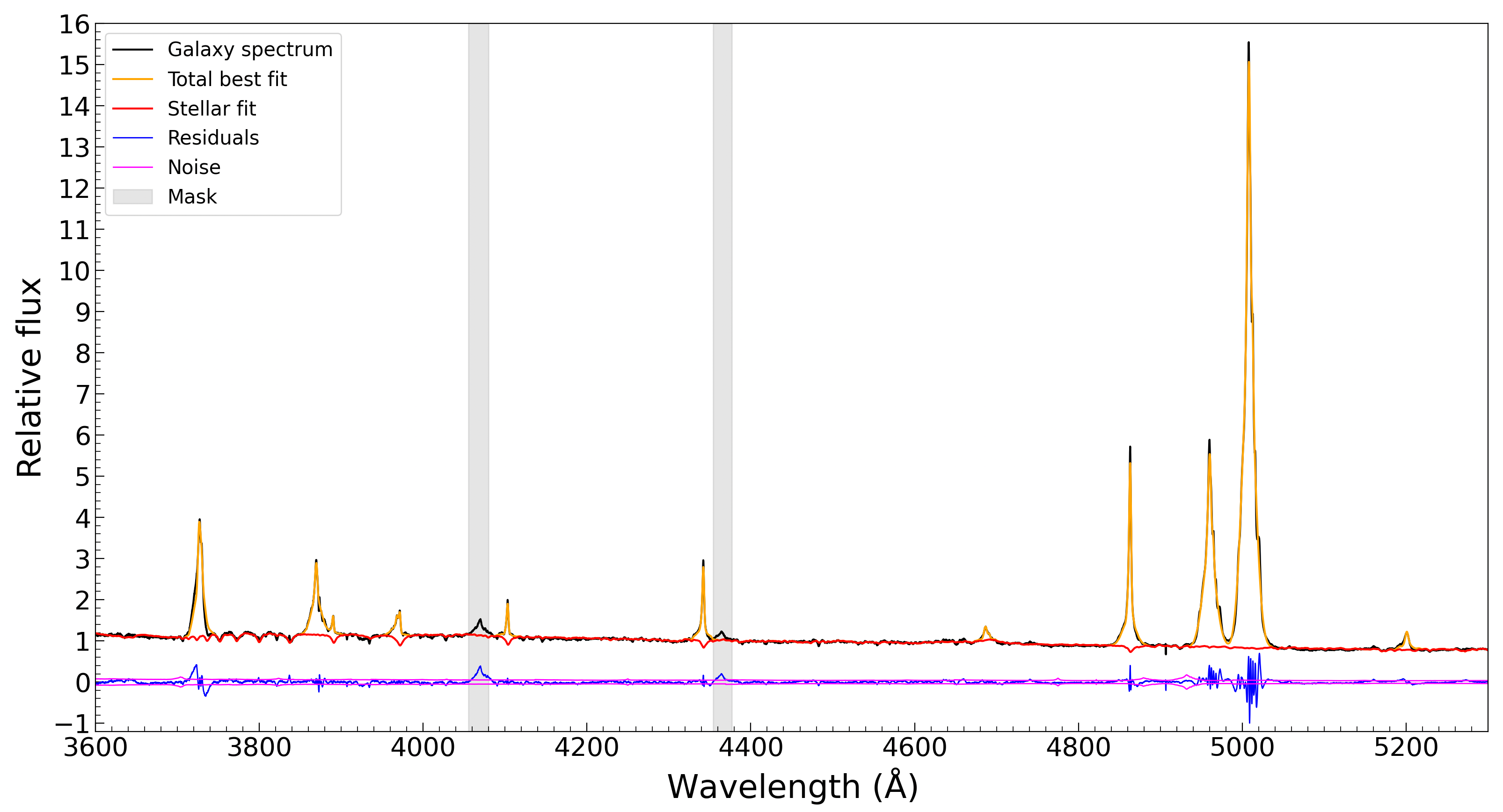}
		\caption{} 
		\label{circle}
	\end{subfigure}
	\begin{subfigure}{\textwidth}
        \centering
		\includegraphics[width=0.72\textwidth]{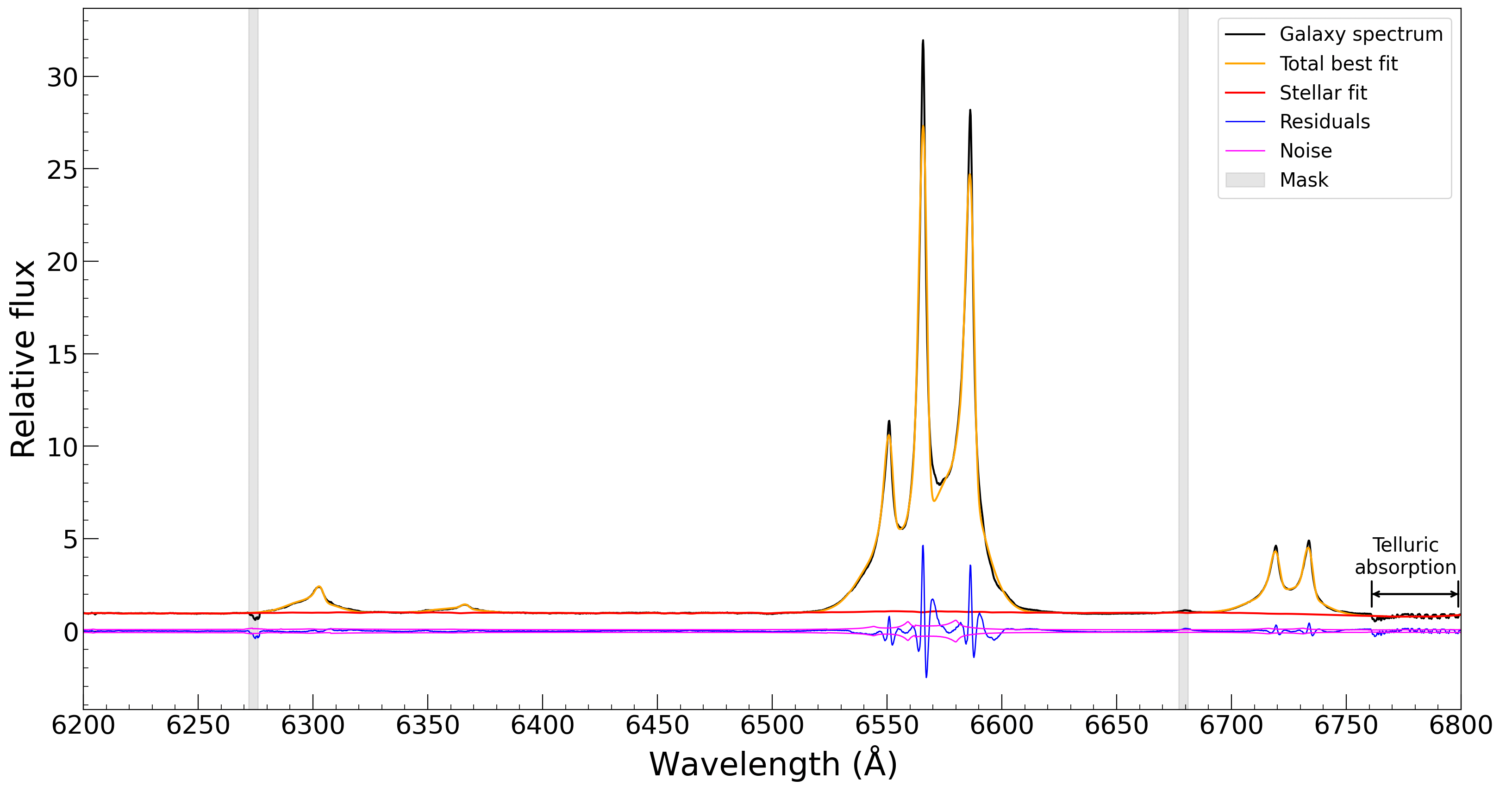}
		\caption{} 
		\label{Mean}
	\end{subfigure}
	\caption{Simultaneously fitting the stellar continuum (red) and emission lines (orange) of the spectrum (black) of galaxy NGC 7130 extracted from the central region in the (a) UVB range and (b) VIS range. Shown below the fit are the residuals (blue) from the \ppxf fitting and the noise spectrum (magenta) from the observation. Emission lines unrelated to the analysis has been masked (grey vertical strips) to avoid interrupting the stellar continuum fitting. Atmospheric absorption does not interfere with the emission lines.} 
    \label{fig:spec_NGC7130}
\end{figure*}

\begin{figure*}
	\includegraphics[width=\columnwidth]{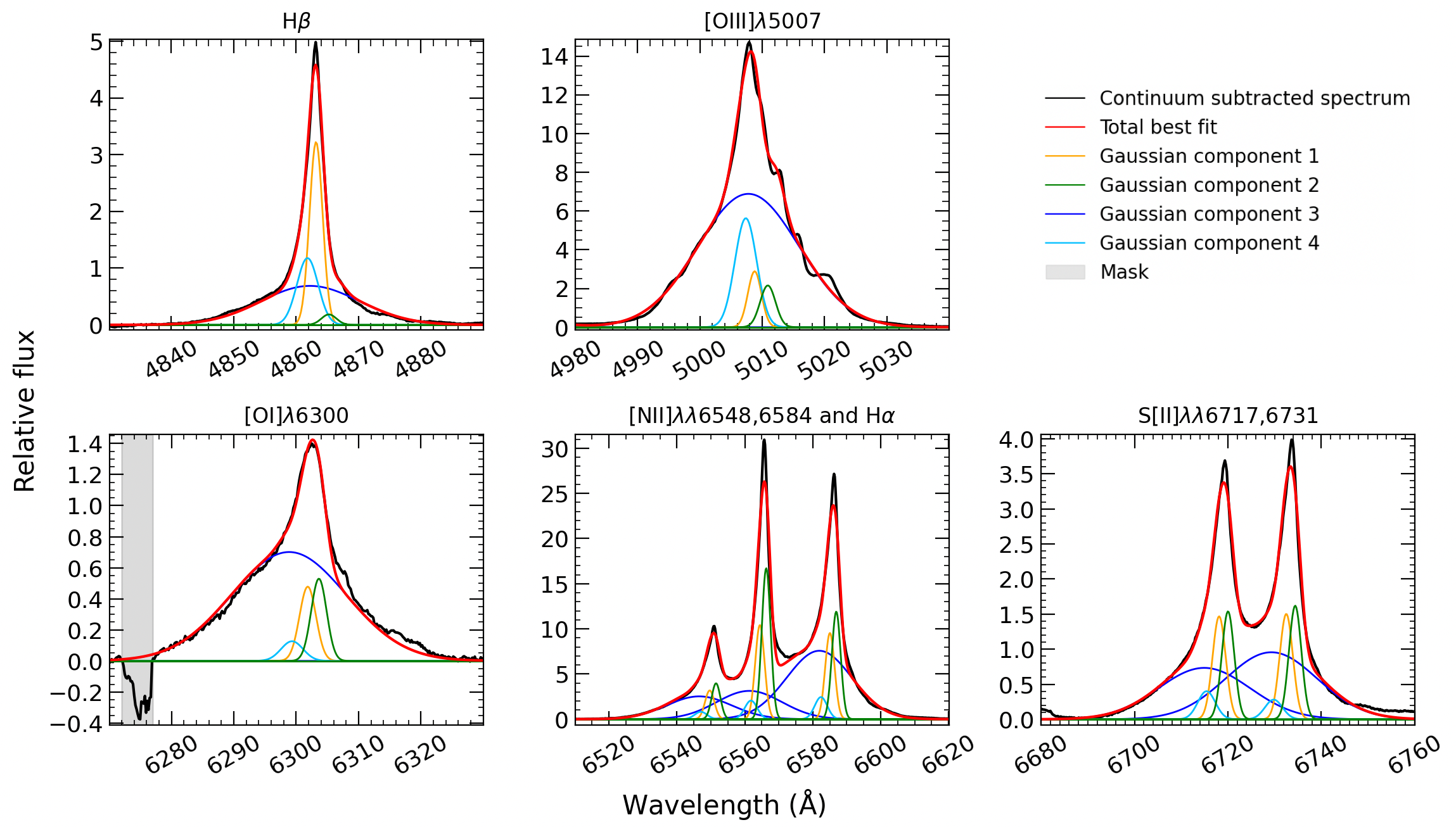}
    \caption{Subsections of continuum-subtracted X-shooter spectra (black) of galaxy NGC 7130, extracted from the central region around the relevant emission lines for BPT-diagnostics, displaying the decomposition by 4 Gaussian components representing the narrow components associated with the gas in the galactic disk and the broad components tracing the outflows.}
    \label{fig:zoom_NGC7130}
\end{figure*}

\begin{figure*}
	\includegraphics[width=\columnwidth]{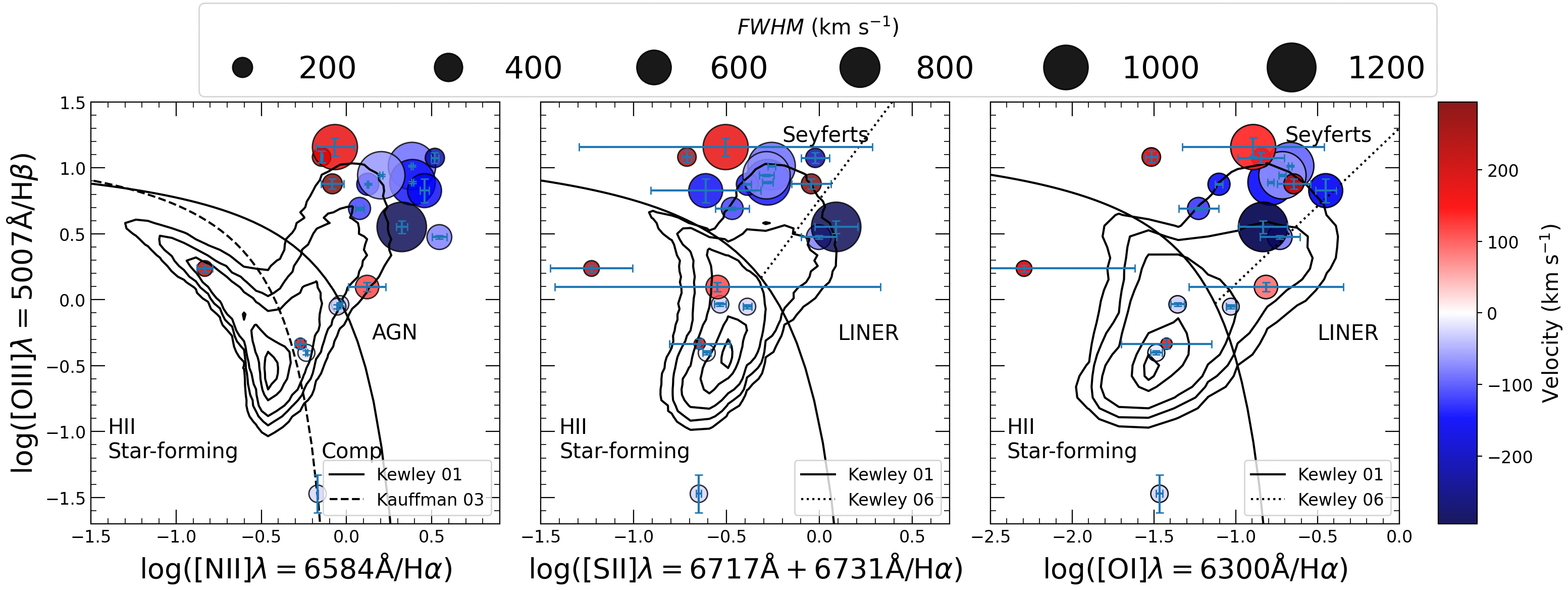}
    \caption{BPT diagnostics diagrams same as~\Cref{fig:NVSSJ151402+015737_bpt}, but for galaxy NGC7130. None of the broad and blueshifted gas components are located in the SF region. Therefore, galaxy NGC 7130 does not host SF in its galactic outflows.}    
    \label{fig:bpt_NGC7130}
\end{figure*}

This section displays the fitting and decomposition of galaxy spectra, along with the BPT diagnostics for the remaining 6 galaxies observed in our ESO X-shooter programme with sufficiently robust outflow detections. IRAS 13120-5453 (see~\Cref{fig:spec_IRAS_13120-5453,fig:zoom_IRAS_13120-5453,fig:bpt_IRAS_13120-5453}) and F13229-2934 (see~\Cref{fig:spec_F13229-2934,fig:zoom_F13229-2934,fig:bpt_F13229-2934}) show indicators of star formation in their BPT diagrams. The star forming components were further investigated with the $R_{23}$ vs $O_{23}$ diagram (see~\Cref{fig:ion}) and we ruled out the possibility of external photoionisation (see~\Cref{section:photoion}). Contributions from shocks have been eliminated as a cause of these SF indicators (see~\Cref{section:shock}). For the rest of the galaxies in our sample, indicators of star formation within galactic outflows have not been observed. Nonetheless, the fitting of spectra and BPT diagnostics of galaxy F22491-1808 (see~\Cref{fig:spec_F22491-1808,fig:zoom_F22491-1808,fig:bpt_F22491-1808}), NVSSJ151402+015737 (see~\Cref{fig:spec_NVSSJ151402+015737,fig:zoom_NVSSJ151402+015737,fig:NVSSJ151402+015737_bpt}), LEDA 166649 (see~\Cref{fig:spec_LEDA166649,fig:zoom_LEDA166649,fig:bpt_LEDA166649}) and NGC 7130 (see~\Cref{fig:spec_NGC7130,fig:zoom_NGC7130,fig:bpt_NGC7130}) are shown.

\onecolumn

\FloatBarrier


\FloatBarrier

\twocolumn
\bsp	
\label{lastpage}
\end{document}